\documentclass[aps,prb,twocolumn,floats]{revtex4}

\usepackage{amssymb}
\usepackage{amsmath}
\usepackage{graphicx}
\usepackage{bm}
\usepackage{mdframed}
\usepackage{color}
\usepackage{gensymb}
\usepackage{comment}
\usepackage{lipsum}
\usepackage{dsfont}
\usepackage{tikz}
\usepackage{appendix}
\usepackage[scr=boondox,  
            cal=esstix]   
           {mathalpha}

\newcommand{\ds}{\displaystyle}
\newcommand{\Cross}{\mathbin{\tikz [x=1.4ex,y=1.4ex,line width=.3ex] \draw (0,0) -- (1,1) (0,1) -- (1,0);}}
\makeatletter
\newcommand*\bigcdot{\mathpalette\bigcdot@{.7}}
\newcommand*\bigcdot@[2]{\mathbin{\vcenter{\hbox{\scalebox{#2}{$\m@th#1\bullet$}}}}}
\makeatother

\begin{document}

\bibliographystyle{prsty}

\title{Transport theory and spin-transfer physics in d-wave altermagnets}

\author{Ricardo Zarzuela$^{1}$, Rodrigo Jaeschke-Ubiergo$^{1}$, Olena Gomonay$^{1}$, Libor \v{S}mejkal$^{1,2,3}$ and Jairo Sinova$^{1,4}$}

\affiliation{$^{1}$Institut f\"{u}r Physik, Johannes Gutenberg Universit\"{a}t Mainz, D-55099 Mainz, Germany\\
$^{2}$Max Planck Institute for the Physics of Complex Systems, N\"{o}thnitzer Str. 38, 01187 Dresden, Germany\\
$^{3}$Institute of Physics, Czech Academy of Sciences, Cukrovarnick\'{a} 10, 162 00, Praha 6, Czech Republic\\
$^{4}$Department of Physics, Texas A\&M University, College Station, Texas, USA
}

\begin{abstract}
{We develop a mesoscale transport theory for the charge and spin degrees of freedom of itinerant carriers in a $d$-wave altermagnet. Our effective Lagrangian description is built upon the slave-boson formulation of the microscopic $t-J$ model. We obtain a spin-polarized diffusive contribution to the effective Hamiltonian, with no counterpart in conventional antiferromagnetism and parametrized by the spin splitting, that is responsible for the so-called spin-splitter effect in $d$-wave altermagnets. We also elucidate the spin-transfer response of the itinerant fluid as well as the spin pumping into the altermagnet, which show previously unidentified combinations of the charge current and spatial partial derivatives (namely, \{$j_{x}^{e}$,$\partial_{y}$\} and \{$j_{y}^{e}$,$\partial_{x}$\}). The emergent spin-transfer physics in $d$-wave altermagnets opens up new possibilities for the dynamics of spin textures, such as the domain-wall motion driven by transverse charge currents. We also consider the effect of elastic distortions in the aforementioned transport properties.}
\end{abstract}
\maketitle

\section{Introduction}

The recent years have witnessed the emergence of altermagnetism, a novel magnetic phase of matter characterized by a vanishing net magnetization and the opposite-spin sublattices being connected only by crystal rotations (possibly combined with a translation or inversion operation) \cite{Smejkal-PRX2022a,Smejkal-PRX2022b,Mazdin-PRX2022,Bai-AFM2024,Jungwirth-arXiv2024}. The resultant electronic band structure exhibits time reversal-symmetry-broken momentum-dependent spin splittings with alternating sign as well as anisotropic even-parity ($d$-, $g$- or $i$-)wave spin-split isoenergy surfaces. Manifestations of these unconventional properties are the crystal Hall effect \cite{Smejkal-SciAdv2020,Mazin-PNAS2021,Feng-NatElectron2022,Reichlova-NComms2024}, the anomalous Nernst effect in compensated organic magnets~\cite{Naka-2019}, the discovery of anisotropic exchanges leading to chiral splitting magnon bands \cite{Smejkal-PRL2023} and the experimental evidence of the spin splitting in altermagnets \cite{Krempasky-Nat2024,Reimers-NCom2024,Lee-PRL2024,Osumi-PRB2024,Yang-arXiv2024,Ding-PRL2024,Zeng-AS2024,Li-arXiv2024,Fedchenko-SciAdv2024}, to name a few. Altermagnets also combine the most prominent features of ferro- and antiferromagnets \cite{Mazdin-PRX2022}, making them promising active elements in future spintronic, spin caloritronic, ultrafast optical, multiferroic and neuromorphic applications \cite{Smejkal-PRX2022b}. In this regard, $d$-wave altermagnets have gathered much interest due to their promising transport properties, epitomized by the spin-splitter effect \cite{Gonzalez-PRL2021,Bose-NatElectron2022}: for instance, the generation of unconventional spin currents with $d$-wave symmetry enables both giant and tunnel magnetoresistance effects to reach 100\% scale \cite{Smejkal-PRX2022c,Shao-NatComms2021}, which makes this class of altermagnets potentially very valuable from the technological standpoint. 

Most of the insight into altermagnetism has hitherto been gained by exploring its physical properties within the band-structure approach. From a mesoscopic standpoint, however, the physics of altermagnets remains largely unexplored. Its understanding can be useful to control and manipulate the stability and dynamics of the emergent (topological) spin textures. For instance, we note the recent successful efforts made in order to build a continuum theory for the exchange interactions in a $d$-wave altermagnet, from which the stability of domain walls and skyrmions was concluded \cite{ Gomonay-Spin2024,Jin-arXiv2024}. 

In this work, we investigate the mesoscale transport of spin and charge in a $d$-wave altermagnetic conductor, with special emphasis on the spin-transfer response of the itinerant fluid as well as the spin-pumping currents generated by the dynamics of spin textures in the magnetic background. We construct an effective Lagrangian for the itinerant carriers based upon the microscopic $t-J$ model for conduction electrons, which is well-known to faithfully describe the electron physics in strongly correlated systems. Remarkably, our formalism predicts the appearance of a spin-polarized diffusive term in the long-wavelength Hamiltonian for itinerant carriers; the corresponding spin-polarized mass tensor exhibits nontrivial off-diagonal elements with polarization parallel to the N\'{e}el order that are parametrized by the $d$-wave altermagnetic spin splitting. These nonzero off-diagonal elements, not present in conventional antiferromagnets, are responsible for the characteristic intertwinement between charge and spin currents in $d$-wave altermagnets, which underpins the spin-splitter effect, namely the generation of transverse time reversal-symmetry-odd nonrelativistic spin-polarized currents \cite{Gonzalez-PRL2021}.

Our effective transport theory also describes the spin-transfer physics of the magnetic system. We have obtained a phenomenological expression for the altermagnetic contribution to the spin-transfer torque, $\bm{\tau}_{\bm{m},\textrm{ST}}^{\textrm{AM}}=\eta_{\textrm{FL}}\big[j^{e}_{x}\partial_{y}+j^{e}_{y}\partial_{x}\big]\bm{n}+\tfrac{\eta_{\textrm{DL}}}{\mathfrak{s}}\bm{n}\Cross\big[j^{e}_{x}\partial_{y}+j^{e}_{y}\partial_{x}\big]\bm{n}$, where the phenomenological constants $\eta_{\textrm{FL}}$ and $\eta_{\textrm{DL}}$ denote the reactive and dissipative projections of the torque, respectively, and are proportional to the $d$-wave spin splitting. Here, $\bm{n}$ denotes the N\'{e}el order parametrizing the magnetic phase. We remark $i$) the distinctive (transverse) combination $j^{e}_{x}\partial_{y}+j^{e}_{y}\partial_{x}$ of the charge current and partial derivatives, which is present in $d$-wave altermagnetic platforms only, and $ii$) the presence of a fieldlike contribution to the spin-transfer torque linear with the N\'{e}el order, which is forbidden in bipartite antiferromagnets. A similar transverse behavior is also obtained for the (Onsager-reciprocal) spin-pumping current, which differs from the \emph{isotropic} dependence expected in conventional antiferromagnetism. This exotic nature of the spin-transfer response has a significant impact on the dynamics of altermagnetic spin textures, such as topological solitons: for instance, charge currents injected transversally to the normal of a domain wall are expected to trigger its dynamics, contrary to the bipartite antiferromagnetic case.

The manuscript is organized as follows. In Section~\ref{Sec2} we introduce and discuss the effective Lagrangian embodying our transport theory for itinerant carriers in the altermagnetic medium, whereas in Sec.~\ref{Sec3} we develop the transport equations for the itinerant charge and spin degrees of freedom, along with the constitutive equations for the corresponding currents. We devote Sec.~\ref{Sec4} to the study of the unconventional spin-transfer and spin-pumping physics in $d$-wave altermagnets and Sec.~\ref{Sec5} to the discussion of the effect of elastic deformations on the altermagnetic features of itinerant transport. We discuss our findings in Sec.~\ref{Discussion} and, in the Appendices, we include the microscopic derivations of our long-wavelength Lagrangian as well as some intermediate identities/derivations.


\begin{figure*}[ht!]
\begin{center}
\includegraphics[width=1.0\textwidth]{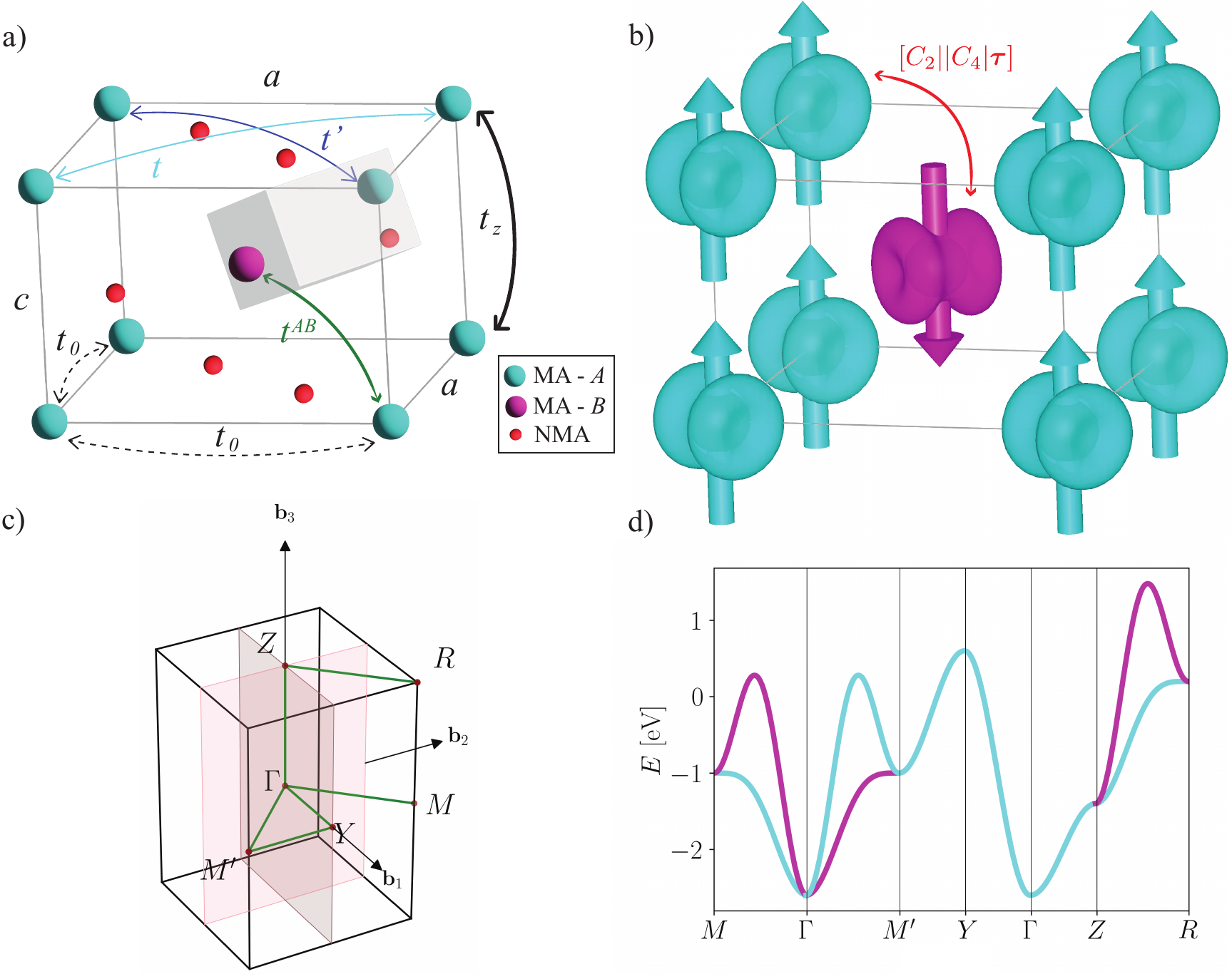}
\caption{(a) Rutile unit cell and tight-binding constants of the model. Turquoise and fuchsia spheres correspond to magnetic atoms (MAs) belonging to sublattices$-A$ and $-B$, respectively. Red spheres depict the nonmagnetic atoms (NMAs) of the crystal structure. The magnetic unit cell, contoured in grey, consists of two atoms belonging to different magnetic sublattices. $t^{AB}$ denotes the interlattice hopping matrix element, whereas $t_{0}, t_{z}, t, t'$ parametrize the intralattice hopping of electrons. (b) Isosurfaces of the $z$-projections $S^{A}_{z}$ (turquoise) and $S^{B}_{z}$ (fuchsia) of the sublattice spin densities centered at the magnetic lattice sites. These opposite-sign isosurfaces are strongly anisotropic, and cannot be mapped into each other by translations or inversions, contrary to the case of conventional antiferromagnets. In particular, due to the $d$-wave character of the altermagnet, the isosurfaces associated to different magnetic sublattices are connected via the spin-group symmetry element $\big[C_{2}||C_{4}|\bm{\tau}\big]$, which, in our model, is embodied in the interchange $AA\, (t,t')\leftrightarrow BB\, (t',t)$ of intralattice tight-binding constants. Arrows depict the net volume-averaged spin at the lattice site. (c) First Brillouin zone of the rutile crystal structure and its high-symmetry points and lines. (d) Band structure generated by the values $t_{0}=-0.2, t_{z}=-0.3, t=-0.5, t'=-0.1$ and $t^{AB}=0$ (in eV) of the tight-binding constants in the Hubbard model~\eqref{eq:Hubbard}. The electronic bands exhibit opposite spin polarizations (as indicated by the same color code of panel (b)) and the band minimum occurs at the $\bm{\Gamma}$-point for this choice of hopping matrix elements.}
\vspace{-0.5cm} 
\label{Fig1}
\end{center}
\end{figure*}

\section{Effective theory of itinerant transport}
\label{Sec2}

We describe the physics of itinerant electrons in the altermagnetic medium by means of the single-band doped $t-J$ model, whose microscopic Hamiltonian reads
\begin{equation}
\label{eq:Hubbard}
H=\sum_{\langle i,j\rangle,\sigma}t_{ij}c_{i,\sigma}^{\dagger}c_{j,\sigma}+\sum_{\langle i,j\rangle}J_{ij}\bm{s}_{i}\bigcdot\bm{s}_{j}.
\end{equation}
The tight-binding term, which describes the electron kinematics within the lattice, and the spin-exchange term are parametrized by the hopping matrix elements $t_{ij}$ and the exchange coupling constants $J_{ij}$ between the lattice sites $\{i,j\}$, respectively. The spin operators are defined as $\bm{s}_{i}=\tfrac{\hbar}{2}\hat{c}_{i,\sigma}\bm{\tau}_{\sigma\sigma'}\hat{c}_{i,\sigma'}$, where $\sigma=\uparrow,\downarrow$ denotes the projection onto the quantization axis of the itinerant spin-$\tfrac{1}{2}$ degree of freedom, and $c_{i,\sigma}$ is the electron annihilation operator at the $i$-th site for the spin projection $\sigma$. We note that, in our notation, bold and $\vec{\,}\,$ represent vectors in the spin and real spaces, respectively, whereas $\bigcdot$ and $\Cross$ denote the scalar and vector products in the spin space, respectively. In what follows, we consider the scenario where the itinerant electrons are embedded into (and flow within) a magnetic background that consists of localized spins interacting via exchange. Furthermore, the rotational symmetry is spontaneously broken in the localized spin sector by the presence of the N\'{e}el order $\bm{n}$, which describes the spin orientation of the ordered phase. 

Our effective theory for itinerant carriers in a collinear $d$-wave altermagnetic conductor will be built upon the continuum limit of the tight-binding Hamiltonian~\eqref{eq:tb_hamiltonian2} and the (path-integral) kinetic Lagrangian~\eqref{eq:kin_term}, which are obtained within the spin-rotation invariant slave-boson framework \cite{Kotliar-PRL1986,Li-PRB1989,Fresard-EPL1991,Fresard-JCM1992}, see Appendix~\ref{AppA1} for a detailed derivation. We note in passing that the spinon degree of freedom parametrizing the itinerant electron in the slave-boson representation adjusts to the sublattice spin densities describing the localized spin background adiabatically, as if $J_{sd}\rightarrow\infty$ in a $s-d$ model description of the spin (itinerant+localized) system. We consider long-wavelength expansions around a set of reference lattice sites of the order-parameter fields describing both the itinerant fluid and the localized spins, namely the holon spinor field $\Psi$ and the sublattice spin densities $\bm{S}^{A}$ and $\bm{S}^{B}$, respectively. Alternatively, we will describe the localized spins via the macroscopic spin density $\bm{m}=\tfrac{1}{2}[\bm{S}_{A}+\bm{S}_{B}]$ and the N\'{e}el order $\bm{n}=\tfrac{1}{2}[\bm{S}_{A}-\bm{S}_{B}]$. We note that the spinor field $\Psi$, which describes the itinerant charge degree of freedom, is not smooth in the long-wavelength regime (in contrast to the sublattice spin densities), since it exhibits an oscillatory behavior at short length scales, characterized by the set $\{\vec{k}_{\nu}\}_{\nu}$ of wavevectors corresponding to the minima of the energy bands. Therefore, in the low-energy long-wavelength limit, we can expand the holon spinor in the following Bloch-type form:  
\begin{equation}
\label{eq:holon_exp}
\Psi(\vec{r})\sim\sum_{\nu}e^{i\vec{k}_{\nu}\cdot\vec{r}}\Psi_{\nu}(\vec{r}),
\end{equation}
where $\{\Psi_{\nu}\}_{\nu}$ are smooth spinor fields over mesoscopic length scales and indexed as well by a \emph{valley} index. It is worth remarking that, in the forthcoming calculations, we will focus on those terms in the effective Lagrangian that are diagonal on the valley indices, since off-diagonal terms will always contribute at the subleading order \cite{FN1}.

The kinetic Lagrangian, rooted in the microscopic path-integral term $\frac{\hbar}{2}\sum_{i,\sigma}\big[\hat{c}_{i,\sigma}^{\dagger}\partial_{0}\hat{c}_{i,\sigma}-\partial_{0}\hat{c}_{i,\sigma}^{\dagger}c_{i,\sigma}\big]$ ($x_{0}=it$ denotes the Wick-rotated time coordinate), is not only diagonal in the valley index, but also independent of it, see Eq.~\eqref{eq:kin_Lagrangian}. The low-energy long-wavelength expansion of the tight-binding term~\eqref{eq:tb_hamiltonian2} is valley dependent, and its general expression will be given and discussed in detail in Appendix~\ref{AppA2}, along with its microscopic derivation for the rutile crystal structure. We will focus hereafter on the valley taking place at the $\mathbf{\Gamma}$-point of the Brillouin zone, see Fig.~\ref{Fig1}(d), which offers the most compact expression for our effective theory without compromising the ensuing physics. The effective Euclidean Lagrangian for holons around the $\bm{\Gamma}$-point reads
\begin{widetext}
\begin{align}
\label{eq:EuclidLag_AM}
\mathcal{L}_{\textrm{eff}}&=\int d\vec{r}\,\bigg[g_{0}(\Psi^{\dagger}\partial_{0}\Psi-\partial_{0}\Psi^{\dagger}\Psi)+g_{0}'\bm{m}\bigcdot(\Psi^{\dagger}\bm{\tau}\partial_{0}\Psi-\partial_{0}\Psi^{\dagger}\bm{\tau}\Psi)+\Phi(\bm{m},\bm{n})\Psi^{\dagger}\Psi+\bm{\omega}(\bm{m},\bm{n})\bigcdot(\Psi^{\dagger}\bm{\tau}\Psi)\\
&\hspace{1.5cm}+\frac{2g_{0}}{\hbar}\vec{\nabla}\Psi^{\dagger}\left[\frac{\hbar^{2}}{2\underline{M}}\right]\vec{\nabla}\Psi+\frac{2g_{0}}{\hbar}\vec{\nabla}\Psi^{\dagger}\left[\frac{\hbar^{2}}{2\underline{\bm{M}}^{s}}\right]\bigcdot\bm{\tau}\vec{\nabla}\Psi+\bm{\mathcal{A}}_{k}\bigcdot(i\Psi^{\dagger}\bm{\tau}\partial_{k}\Psi-i\partial_{k}\Psi^{\dagger}\bm{\tau}\Psi)\bigg].\nonumber
\end{align}
\end{widetext}
We note that this continuum theory can also be built upon symmetry grounds and has been expanded up to second order in the macroscopic spin density and the partial derivatives of the Fermi field and the N\'{e}el order. Furthermore, we have disregarded in the above equation the valley subindex $\Gamma$ for the sake of notational simplicity. Here, $[\underline{M}^{-1}]$ and $[(\underline{\bm{M}}^{s})^{-1}]$ denote an effective mass tensor and an effective spin-polarized mass tensor for the conduction band, respectively, $\bm{\mathcal{A}}_{k}$ are the spatial components of the emergent spin-textured gauge field, and $k=x,y,z$ runs over spatial indices. 

There is a clear physical interpretation of the different terms arising in the effective Hamiltonian~\eqref{eq:EuclidLag_AM}: from left to right, the first one has an electrostatic origin, since it couples the carrier density $\rho=\Psi^{\dagger}\Psi$ to an emergent spin-textured electric potential $\Phi({\bm{m},\bm{n}})\equiv g_{1}^{m}\bm{m}^{2}+g_{1}^{z}\,\partial_{z}\bm{n}\bigcdot\partial_{z}\bm{n}+g_{1}^{xy}\big(\vec{\nabla}_{xy}\bm{n}\big)^{2}+g_{1}^{\textrm{AM}}\big(\partial_{x}\bm{m}\bigcdot\partial_{y}\bm{n}+\partial_{y}\bm{m}\bigcdot\partial_{x}\bm{n}\big)$. The second term, embodying a Zeeman-like coupling for the itinerant spin density $\bm{s}=\tfrac{\hbar}{2}\Psi^{\dagger}\bm{\tau}\Psi$, is responsible for the precessional spin dynamics of the itinerant charge fluid. The corresponding spin precession vector $\bm{\omega}(\bm{m},\bm{n})\equiv g_{0}''\bm{n}\Cross\partial_{t}\bm{n}+g_{2}^{m}\bm{m}+g_{2}^{\textrm{AM}}\partial_{xy}\bm{n}$ presents a (dynamical) Coriolis-type contribution proportional to the order-parameter angular velocity, which originates in the fact that the itinerant charge flows in a rotating spin frame of reference adjusted to the localized spin background. The other two contributions have a spin-exchange origin and are proportional to the total background magnetization and the second order derivative of the N\'{e}el order. The third term is the usual kinetic term parametrized by the effective mass tensor $\big[\underline{M}^{-1}\big]$ of the carriers, which is responsible for the diffusive contributions to both charge and spin currents. The fourth term describes a spin-polarized diffusion of the itinerant carriers, which is parametrized by an effective spin-polarized mass tensor $[(\underline{\bm{M}}^{s})^{-1}]$. The fifth term describes an emergent coupling between the itinerant spin current and the emergent non-Abelian gauge fields
\begin{align}
\label{eq:gauge_fields}
&\bm{\mathcal{A}}_{x}\equiv g_{4}^{xy}\,\bm{n}\Cross\partial_{x}\bm{n}+g_{4}^{\textrm{AM}}\,\bm{m}\Cross\partial_{y}\bm{n},\\
&\bm{\mathcal{A}}_{y}\equiv g_{4}^{xy}\,\bm{n}\Cross\partial_{y}\bm{n}+g_{4}^{\textrm{AM}}\,\bm{m}\Cross\partial_{x}\bm{n},\nonumber\\
&\bm{\mathcal{A}}_{z}\equiv g_{4}^{z}\,\bm{n}\Cross\partial_{z}\bm{n}.\nonumber
\end{align}
Remarkably, as we will show and discuss in the next sections, this last term captures the essential features of the spin-transfer physics observed in collinear altermagnetic platforms, since the motion of the itinerant charge fluid favors a twist of the order parameter and viceversa \cite{Shraiman-PRL1988}. Furthermore, this coupling is also responsible for the topological Hall response of the magnetic medium. The effective mass tensor is diagonal and takes the form $[\underline{M}^{-1}]=\textrm{diag}(m_{\parallel}^{-1},m_{\parallel}^{-1},m_{z}^{-1})$ for the rutile crystal structure. On the contrary, the spin-polarized mass tensor exhibits nonzero off-diagonal elements within the $xy$ plane:
\begin{equation}
\label{eq:sp_mass_tensor}
\big[(\underline{\bm{M}}^{s})^{-1}\big]=
\begin{pmatrix}
-\frac{1}{m^{s}_{\parallel}}\bm{m} & \frac{1}{m^{s}_{xy}}\bm{n} & \bm{0}\\
\frac{1}{m^{s}_{xy}}\bm{n} & -\frac{1}{m^{s}_{\parallel}}\bm{m} & \bm{0}\\
\bm{0} & \bm{0} & -\frac{1}{m^{s}_{z}}\bm{m}
\end{pmatrix}.
\end{equation}
Its diagonal elements are proportional to the macroscopic spin density of the magnet and, therefore, will contribute at the subleading order to the holon conduction. However, its nonzero off-diagonal elements, being parametrized by the N\'{e}el order, will contribute at the leading order.  The explicit expression of the phenomenological constants $g_{0}$'s, $g_{1}$'s, $g_{2}$'s, $g_{4}$'s, $m_{\parallel}$, $m_{z}$, $m_{\parallel}^{s}$, $m_{z}^{s}$ and $m_{xy}^{s}$ in terms of the microscopic parameters of the rutile model will be provided in the Appendix~\ref{AppA2}. As we will show in the next Sections, these off-diagonal elements will be responsible for the spin-splitter effect in $d$-wave altermagnets.

\section{Hydrodynamic theory and the spin-splitter effect}
\label{Sec3}

Transport equations for the itinerant charge and spin degrees of freedom can be obtained by exploiting the hydrodynamic properties of nonrelativistic SU(2) Yang-Mills theories \cite{Jin-JPA2006}. Our starting point are the saddle-point equations for the Fermi field and its complex conjugate:
\begin{widetext}
\begin{align}
\label{eq:sadpoint_psi}
\frac{\delta\mathcal{L}_{\textrm{eff}}}{\delta\Psi^{\dagger}}&=2g_{0}\partial_{0}\Psi+g_{0}'(\bm{m}\bigcdot\bm{\tau})\partial_{0}\Psi+g_{0}'\partial_{0}(\bm{m}\bigcdot\bm{\tau}\Psi)+\Phi\Psi+(\bm{\omega}\bigcdot\bm{\tau})\Psi-\tfrac{2g_{0}}{\hbar}\partial_{\alpha}\big(\big[\tfrac{\hbar^{2}}{2\underline{M}}\big]_{\alpha\beta}\partial_{\beta}\Psi\big)\\
&\hspace{0.5cm}-\tfrac{2g_{0}}{\hbar}\partial_{\alpha}\big(\big[\tfrac{\hbar^{2}}{2\underline{\bm{M}}^{s}}\bigcdot\bm{\tau}\big]_{\alpha\beta}\partial_{\beta}\Psi\big)+i(\bm{\mathcal{A}}_{\alpha}\bigcdot\bm{\tau})\partial_{\alpha}\Psi+i\partial_{\alpha}(\bm{\mathcal{A}}_{\alpha}\bigcdot\bm{\tau}\Psi)=0,\nonumber\\
\label{eq:sadpoint_psidag}
\frac{\delta\mathcal{L}_{\textrm{eff}}}{\delta\Psi}&=-2g_{0}\partial_{0}\Psi^{\dagger}-g_{0}'\partial_{0}(\Psi^{\dagger}\bm{m}\bigcdot\bm{\tau})-g_{0}'\partial_{0}\Psi^{\dagger}(\bm{m}\bigcdot\bm{\tau})+\Phi\Psi^{\dagger}+\Psi^{\dagger}(\bm{\omega}\bigcdot\bm{\tau})-\tfrac{2g_{0}}{\hbar}\partial_{\beta}\big(\partial_{\alpha}\Psi^{\dagger}\big[\tfrac{\hbar^{2}}{2\underline{M}}\big]_{\alpha\beta}\big)\\
&\hspace{0.5cm}-\tfrac{2g_{0}}{\hbar}\partial_{\beta}\big(\partial_{\alpha}\Psi^{\dagger}\big[\tfrac{\hbar^{2}}{2\underline{\bm{M}}^{s}}\bigcdot\bm{\tau}\big]_{\alpha\beta}\big)-i(\bm{\mathcal{A}}_{\alpha}\bigcdot\partial_{\alpha}\Psi^{\dagger}\bm{\tau})-i\partial_{\alpha}(\bm{\mathcal{A}}_{\alpha}\bigcdot\Psi^{\dagger}\bm{\tau})=0,\nonumber
\end{align}
\end{widetext}
which, via the linear combination $\Psi^{\dagger}\textrm{Eq.}~\eqref{eq:sadpoint_psi}-\textrm{Eq.}~\eqref{eq:sadpoint_psidag}\Psi$, yield the following hydrodynamic equation for the probability density:
\begin{equation}
\label{eq:hydro_eq_prob}
\partial_{t}\rho+\partial_{\alpha}j_{\alpha}=-\tfrac{2g_{0}'}{\hbar g_{0}}\partial_{t}[\bm{m}\bigcdot\bm{s}].
\end{equation}
Here, the spatial components of the probability current are given by
\begin{widetext}
\begin{align}
\label{eq:const_rel_p_curr}
j_{x}=&\frac{\hbar}{2m_{\parallel}}\big(i\partial_{x}\Psi^{\dagger}\Psi-i\Psi^{\dagger}\partial_{x}\Psi\big)-\frac{2}{\hbar g_{0}}\bm{\mathcal{A}}_{x}\bigcdot\bm{s}+\frac{\hbar}{2m^{s}_{xy}}\bm{n}\bigcdot\big(i\partial_{y}\Psi^{\dagger}\bm{\tau}\Psi-i\Psi^{\dagger}\bm{\tau}\partial_{y}\Psi\big),\\
j_{y}=&\frac{\hbar}{2m_{\parallel}}\big(i\partial_{y}\Psi^{\dagger}\Psi-i\Psi^{\dagger}\partial_{y}\Psi\big)-\frac{2}{\hbar g_{0}}\bm{\mathcal{A}}_{y}\bigcdot\bm{s}+\frac{\hbar}{2m^{s}_{xy}}\bm{n}\bigcdot\big(i\partial_{x}\Psi^{\dagger}\bm{\tau}\Psi-i\Psi^{\dagger}\bm{\tau}\partial_{x}\Psi\big),\nonumber\\
j_{z}=&\frac{\hbar}{2m_{z}}\big(i\partial_{z}\Psi^{\dagger}\Psi-i\Psi^{\dagger}\partial_{z}\Psi\big)-\frac{2}{\hbar g_{0}}\bm{\mathcal{A}}_{z}\bigcdot\bm{s},\nonumber
\end{align}
\end{widetext}
where we have omitted diffusive terms linear with the total magnetization, which contribute at the subleading order. We note the presence of a source term in the right-hand side of Eq.~\eqref{eq:hydro_eq_prob}, rooted in the time evolution of total magnetization vector projected onto the itinerant spin density. It can be disregarded in the case of altermagnets since the macroscopic spin density represents a secondary order parameter of negligible magnitude in the regime of low-frequency dynamics and moderate external magnetic fields. Therefore, Eq.~\eqref{eq:hydro_eq_prob} yields the continuity equation for the itinerant charge.

The transport equation for the itinerant spin density is obtained by combining the above saddle-point equations in the form $\big(\Psi^{\dagger}\frac{\hbar}{2}\bm{\tau}\big)\textrm{Eq.}~\eqref{eq:sadpoint_psi}-\textrm{Eq.}~\eqref{eq:sadpoint_psidag}\big(\frac{\hbar}{2}\bm{\tau}\hspace{0.05cm}\Psi\big)$, and reads
\begin{widetext}
\begin{align}
\label{eq:hydro_eq_spin}
\partial_{t}\bm{s}+\partial_{\alpha}\bm{J}_{\alpha}=&-\tfrac{g_{0}'}{g_{0}}\partial_{t}\big[\tfrac{\hbar}{2}\bm{m}\rho\big]+\tfrac{g_{0}'}{g_{0}}\bm{m}\Cross\big(i\partial_{t}\Psi^{\dagger}\tfrac{\hbar}{2}\bm{\tau}\Psi-i\Psi^{\dagger}\tfrac{\hbar}{2}\bm{\tau}\partial_{t}\Psi\big)+\tfrac{1}{g_{0}}\bm{\omega}\Cross\bm{s}+\left[\frac{\hbar}{\underline{\bm{M}}^{s}}\right]_{\alpha\beta}\Cross\big(\partial_{\alpha}\Psi^{\dagger}\tfrac{\hbar}{2}\bm{\tau}\partial_{\beta}\Psi\big)\\
&\hspace{0.5cm}-\tfrac{1}{g_{0}}\bm{\mathcal{A}}_{\alpha}\Cross\big(i\partial_{\alpha}\Psi^{\dagger}\tfrac{\hbar}{2}\bm{\tau}\Psi-i\Psi^{\dagger}\tfrac{\hbar}{2}\bm{\tau}\partial_{\alpha}\Psi\big).\nonumber
\end{align}
\end{widetext}
This dynamical equation must be complemented with the following constitutive relations for the spin current:
\begin{widetext}
\begin{align}
\label{eq:const_real_s_curr}
\bm{J}_{x}=&\frac{\hbar^{2}}{4m_{\parallel}}\big(i\partial_{x}\Psi^{\dagger}\bm{\tau}\Psi-i\Psi^{\dagger}\bm{\tau}\partial_{x}\Psi\big)-\frac{\hbar}{2g_{0}}\rho\,\bm{\mathcal{A}}_{x}+\frac{\hbar}{2m^{s}_{xy}}\left[\frac{\hbar}{2}\bm{n}\big(i\partial_{y}\Psi^{\dagger}\Psi-i\Psi^{\dagger}\partial_{y}\Psi\big)+\bm{n}\Cross\partial_{y}\bm{s}\right],\\
\bm{J}_{y}=&\frac{\hbar^{2}}{4m_{\parallel}}\big(i\partial_{y}\Psi^{\dagger}\bm{\tau}\Psi-i\Psi^{\dagger}\bm{\tau}\partial_{y}\Psi\big)-\frac{\hbar}{2g_{0}}\rho\,\bm{\mathcal{A}}_{y}+\frac{\hbar}{2m^{s}_{xy}}\left[\frac{\hbar}{2}\bm{n}\big(i\partial_{x}\Psi^{\dagger}\Psi-i\Psi^{\dagger}\partial_{x}\Psi\big)+\bm{n}\Cross\partial_{x}\bm{s}\right],\nonumber\\
\bm{J}_{z}=&\frac{\hbar^{2}}{4m_{z}}\big(i\partial_{z}\Psi^{\dagger}\bm{\tau}\Psi-i\Psi^{\dagger}\bm{\tau}\partial_{z}\Psi\big)-\frac{\hbar}{2g_{0}}\rho\,\bm{\mathcal{A}}_{z}.\nonumber
\end{align}
\end{widetext}
Once again, we have omitted diffusive terms linear with the total magnetization, which contribute at the subleading order. We note the presence of source terms in Eq.~\eqref{eq:hydro_eq_spin}, which act as magnetic torques on the itinerant spins: the first term embodies the time variation of an adiabatically-generated itinerant spin density, understood as the spin density corresponding to the itinerant carriers if those followed the (background) total magnetization adiabatically. The second magnetic torque emerges from the coupling between the macroscopic spin density and the real-timelike spin current. The third term describes the usual spin precession under the action of the effective (magnetic) field $\bm{\omega}$. The fourth magnetic torque stems from the spin-filtering diffusive term in the effective Lagrangian~\eqref{eq:EuclidLag_AM} and, last but not least, the fifth term originates in the aforementioned coupling between the non-Abelian gauge field and the spatial components of the itinerant spin current.

Remarkably, in altermagnetic platforms the dynamics of both itinerant charge and spin densities are intertwined: more specifically, the spin-filtering diffusive term of the effective Euclidean Lagrangian couples both charge and spin currents in the aforementioned constitutive equations. This charge current-spin current coupling is parametrized by the spin-polarized effective mass $m_{xy}^{s}$ and, as we will show next, is responsible for the spin-splitter effect observed in $d$-wave altermagnets such as RuO$_{2}$~\cite{Bose-NatElectron2022}: we will assume in what follows a uniform magnetic background (order parameter) as well as a uniform itinerant spin density, so that we can solely focus on the interplay between the diffusive contributions to both charge and spin currents. By combining the expressions for the $x$ and $y$ components of Eqs.~\eqref{eq:const_rel_p_curr} and~\eqref{eq:const_real_s_curr}, we obtain the compact expression
\begin{equation}
\label{eq:spin_splitter}
\bm{J}_{\alpha}=\frac{\hbar^{2}}{4m_{\parallel}}\mathcal{P}\big(i\partial_{\alpha}\Psi^{\dagger}\bm{\tau}\Psi-i\Psi^{\dagger}\bm{\tau}\partial_{\alpha}\Psi\big)+\frac{m_{\parallel}}{m_{xy}^{s}}\sigma_{x}|_{\alpha\beta}\frac{\hbar}{2}\bm{n}\,j_{\beta},
\end{equation}
where $\mathcal{P}\equiv \mathds{1}-(m_{\parallel}/m_{xy}^{s})^{2}\,\bm{n}[\bm{n}\,\bigcdot\,]$ is a spin-space projector operator, $\alpha,\beta=x,y$ and $\sigma_{x}$ is the Pauli $x$ matrix. Therefore, the in-plane components of the spin current split into two contributions: the first one (up to the projection $\mathcal{P}$) is given by the usual quantum-mechanical diffusive term for the spin current, so that it represents its \emph{isotropic} contribution. The second one, however, is polarized along the direction of the N\'{e}el order and is driven by the transverse component of the charge current, therefore representing the \emph{d-wave} contribution to the spin current. 

Further insight into the ensuing transport physics can be gained by studying thoroughly the first diffusive term. We note here that, in the scenario where altermagnetism is switched off (namely, $t=t'$), the magnetic system becomes a conventional bipartite antiferromagnet. Therefore, Eq.~\eqref{eq:spin_splitter} should reduce to the well-known expression for the spin-current flowing within an antiferromagnetic conductor in the limit $1/m_{xy}^{s}\rightarrow0$. Let $\bm{n}=\mathfrak{s}(\sin\theta\cos\phi,\sin\theta\sin\phi,\cos\theta)$ be the spherical parametrization of the $s$-wave projection of the N\'{e}el order. Then, the $s$-wave projection of the macroscopic spin density can be cast as $\bm{m}=m_{\phi}(-\sin\phi,\cos\phi,0)+m_{\theta}(-\cos\theta\cos\phi,-\cos\theta\sin\phi,\sin\theta)\equiv\bm{m}_{\phi}+\bm{m}_{\theta}$, where $m_{\phi},m_{\theta}$ are real parameters. Note that, by definition, $\bm{m}\bigcdot\bm{n}=0$. The itinerant (holon) fluid is in a mixed ensemble described by the density matrix $\hat{\rho}=p_{\phi}\big|\Psi_{\phi}\rangle\langle\Psi_{\phi}\big|+p_{\theta}\big|\Psi_{\theta}\rangle\langle\Psi_{\theta}\big|$, with pure states given by
\begin{align}
\label{eq:pure_states}
\Psi_{\phi}&=\sqrt{\frac{m_{\phi}}{2}}\begin{pmatrix}
-ie^{-i\phi}\\
1
\end{pmatrix}e^{i\alpha_{\phi}},\\
\Psi_{\theta}&=\sqrt{\frac{m_{\theta}}{2}}\begin{pmatrix}
e^{-i\phi}\cos\tfrac{\theta}{2}+e^{-i\phi}\sin\tfrac{\theta}{2}\\
\sin\tfrac{\theta}{2}-\cos\tfrac{\theta}{2}
\end{pmatrix}e^{i\alpha_{\theta}}.\nonumber
\end{align}
We note that the holon wave phases $\alpha_{\phi},\alpha_{\theta}$ satisfy the saddle point equation~\eqref{eq:sadpoint_psi} for a uniform magnetic background. By introducing the current operators $\hat{\vec{j}}_{l}=\tfrac{1}{m_{\parallel}}\hat{\vec{p}}$ (probability) and $\hat{\vec{\bm{J}}}_{l}=\tfrac{1}{2}\big[\hat{\vec{j}}_{l}\tfrac{\hbar}{2}\bm{\tau}+\tfrac{\hbar}{2}\bm{\tau}\hat{\vec{j}}_{l}\big]$ (spin), we obtain the following expression for the first term in Eq.~\eqref{eq:spin_splitter}:
\begin{align}
\label{eq:spin_current}
\vec{\bm{J}}_{l}&=\textrm{Tr}\big[\hat{\rho}\hat{\vec{\bm{J}}}_{l}\big]=p_{\phi}\langle\Psi_{\phi}\big|\hat{\vec{\bm{J}}}_{l}\big|\Psi_{\phi}\rangle+p_{\theta}\langle\Psi_{\theta}\big|\hat{\vec{\bm{J}}}_{l}\big|\Psi_{\theta}\rangle\\
&=\tfrac{\hbar^{2}}{4m_{\parallel}}\big[p_{\phi}\bm{m}_{\phi}\big(i\vec{\nabla}\alpha_{\phi}^{*}\alpha_{\phi}-i\alpha_{\phi}^{*}\vec{\nabla}\alpha_{\phi}\big)\nonumber\\
&\hspace{1.5cm}+p_{\theta}\bm{m}_{\theta}\big(i\vec{\nabla}\alpha_{\theta}^{*}\alpha_{\theta}-i\alpha_{\theta}^{*}\vec{\nabla}\alpha_{\theta}\big)\big],\nonumber
\end{align}
where we have accounted for the fact that $\mathcal{P}\equiv\mathds{1}$ in this case since $\bm{n}\perp \bm{m}_{\phi,\theta}$. In the absence of external magnetic fields and spin-orbit interactions, the localized spin background is magnetically compensated, namely $\bm{m}_{\phi}=\bm{m}_{\theta}=\bm{0}$. As a result, the above two contributions to the isotropic spin current, which are polarized along the $\bm{m}_{\phi}$ and $\bm{m}_{\theta}$ directions, respectively, vanish identically. We can therefore conclude that the total spin current flowing within the altermagnetic conductor reads
\begin{equation}
\label{eq:spin_splitter2}
\begin{pmatrix}
\bm{J}_{x}\\
\bm{J}_{y}
\end{pmatrix}=
\begin{pmatrix}
\bm{0}& \bm{\sigma}_{xy}\\
\bm{\sigma}_{xy} &\bm{0}
\end{pmatrix}
\begin{pmatrix}
j_{x}\\
j_{y}
\end{pmatrix}.
\end{equation} 
By assuming that the electric conductivity tensor in the nonrelativistic regime is isotropic in the $xy$ plane \cite{FN2}, $e\vec{j}\equiv\vartheta\vec{E}$, with $\vec{E}$ being the driving electric field, the spin conductivity tensor is off-diagonal with components $\bm{\sigma}^{s}_{xy}\equiv\tfrac{\vartheta}{e}\bm{\sigma}_{xy}=\tfrac{m_{\parallel}}{m_{xy}^{s}}\vartheta\tfrac{\hbar}{2e}\bm{n}$, which is the real-space realization of the spin-splitter effect. It is worth remarking again that, when altermagnetism is switched off (namely, $t=t'$), we obtain $\vec{\bm{J}}=\vec{\bm{0}}$. The absence/vanishing of spin currents is what one would expect in the antiferromagnetic scenario in the absence of spin-orbit interactions. 

\section{Spin-transfer and spin-pumping physics}
\label{Sec4}

\subsection{Spin-transfer torques}

We proceed next to elucidate the response, at the mean-field level, of the itinerant charge fluid to the presence of a magnetic background. We note that the thermodynamic description of the itinerant carriers is achieved by means of the thermodynamic variables $\{\rho,\bm{s}\}$ and their conjugated thermodynamic fluxes $\{\vec{j},\vec{\bm{J}}\}$. With account of the identity~\eqref{eq:ident2}, the Hamiltonian term of the Euclidean Lagrangian can be recast as the free energy
\begin{widetext}
\begin{align}
\label{eq:energy_itinerant}
\mathcal{E}_{\textrm{it}}[\Psi,\bm{n},\bm{m}]&=\int d\vec{r}\,\Bigg[\Phi\rho+\tfrac{2}{\hbar}\bm{\omega}\bigcdot\bm{s}-\tfrac{2m_{z}}{\hbar}\Big(\tfrac{2}{\hbar}\bm{J}_{z}\bigcdot\bm{\mathcal{A}}_{z}+\tfrac{1}{g_{0}}\rho\bm{\mathcal{A}}_{z}^{2}\Big)-\tfrac{2m_{\parallel}}{\hbar}\sum_{\kappa=x,y}\Big(\tfrac{2}{\hbar}\bm{J}_{\kappa}\bigcdot\bm{\mathcal{A}}_{\kappa}+\tfrac{1}{g_{0}}\rho\bm{\mathcal{A}}_{\kappa}^{2}\\
&+\tfrac{1}{(m_{xy}^{s}/m_{\parallel})^{2}-\mathfrak{s}^{2}}\Big[\tfrac{2}{\hbar}(\bm{n}\bigcdot\bm{\mathcal{A}}_{\kappa})(\bm{n}\bigcdot\bm{J}_{\kappa})+\tfrac{\rho}{g_{0}}(\bm{n}\bigcdot\bm{\mathcal{A}}_{\kappa})^{2}\Big]\Big)+\tfrac{2m_{\parallel}}{\hbar m_{xy}^{s}}\sum_{\kappa,\beta=x,y}\Big[\sigma_{x}|_{\kappa\beta}\bm{\mathcal{A}}_{\kappa}\bigcdot(\bm{n}\Cross\partial_{\beta}\bm{s})\Big]\nonumber\\
&+\tfrac{2}{\hbar}\tfrac{m_{xy}^{s}}{(m_{xy}^{s}/m_{\parallel})^{2}-\mathfrak{s}^{2}}\sum_{\kappa,\beta=x,y}\Big[\sigma_{x}|_{\kappa\beta}\big(j_{\beta}+\tfrac{2}{\hbar g_{0}}\bm{\mathcal{A}}_{\beta}\bigcdot\bm{s}\big)(\bm{n}\bigcdot\bm{\mathcal{A}}_{\kappa})\Big]+\tfrac{2g_{0}}{\hbar}\vec{\nabla}\Psi^{\dagger}\left[\tfrac{\hbar^{2}}{2\underline{M}}\right]\vec{\nabla}\Psi+\tfrac{2g_{0}}{\hbar}\vec{\nabla}\Psi^{\dagger}\left[\tfrac{\hbar^{2}}{2\underline{\bm{M}}^{s}}\right]\bigcdot\bm{\tau}\vec{\nabla}\Psi\Bigg].\nonumber
\end{align}
\end{widetext}
The lattice of localized spins is described by the following minimal energy model for a $d$-wave altermagnet, $\ds \mathcal{E}_{d-\textrm{wave}}[\bm{m},\bm{n}]=\int d\vec{r}\,\Big[\tfrac{\bm{m}^{2}}{2\chi_{0}}+A_{z,0}\partial_{z}\bm{n}\bigcdot\partial_{z}\bm{n}+A_{xy,0}\times$ $(\vec{\nabla}_{xy}\bm{n})^{2}+A_{\textrm{AM},0}(\partial_{x}\bm{m}\bigcdot\partial_{y}\bm{n}+\partial_{y}\bm{m}\bigcdot\partial_{x}\bm{n})\Big]$, where $\chi_{0}$ and $A_{z,0},A_{xy,0},A_{\textrm{AM},0}$ are the bare spin susceptibility and spin stiffness constants of the magnet, respectively \cite{Gomonay-npjSpin2024}. The total free energy of the collinear altermagnetic conductor results from the addition of these two contributions, $\mathcal{E}[\Psi,\bm{n},\bm{m}]=\mathcal{E}_{\textrm{it}}[\Psi,\bm{n},\bm{m}]+\mathcal{E}_{d-\textrm{wave}}[\bm{n},\bm{m}]$. In particular, the interaction with the itinerant degrees of freedom yields, through the electrostatic potential term, the renormalization of the phenomenological constants of the minimal energy model for $d$-wave altermagnets:
\begin{align}
\label{eq:renorm_ctes}
&\frac{1}{\chi}=\frac{1}{\chi_{0}}+2g_{1}^{m}\rho,\hspace{0.3cm}A_{z}=A_{z,0}+g_{1}^{z}\rho,\\
&A_{xy}=A_{xy,0}+g_{1}^{xy}\rho,\hspace{0.3cm}A_{\textrm{AM}}=A_{\textrm{AM},0}+g_{1}^{\textrm{AM}}\rho.\nonumber
\end{align}

The reactive (fieldlike) spin torque acting upon the localized magnetic background of a $d$-wave altermagnet can be obtained via the Poisson-Liouville dynamical equation $\partial_{t}\bm{m}=\big\{\bm{m},\mathcal{E}\big\}$ within the Hamilton symplectic framework, where $\{\cdot,\cdot\}$ denotes the Poisson brackets of the system. We will focus hereafter on the coupling term $\mathcal{E}_{\textrm{ST}}=-\tfrac{4m_{\parallel}}{\hbar^{2}}\big[\bm{J}_{x}\bigcdot\bm{\mathcal{A}}_{x}+\bm{J}_{y}\bigcdot\bm{\mathcal{A}}_{y}\big]$ in the energy functional, which is the one describing/responsible for the altermagnetic spin-transfer response of the itinerant fluid. The Poisson bracket structure describing the localized magnetic background is given by $\big\{m^{i}(\vec{r}),m^{j}(\vec{r}\,')\big\}=\epsilon_{ijk}m^{k}(\vec{r})\delta(\vec{r}-\vec{r}\,')$, $\big\{m^{i}(\vec{r}),n^{j}(\vec{r}\,')\big\}=\epsilon_{ijk}n^{k}(\vec{r})\delta(\vec{r}-\vec{r}\,')$ and $\big\{n^{i}(\vec{r}),n^{j}(\vec{r}\,')\big\}=0$, whereas the functional derivatives of $\mathcal{E}_{\textrm{ST}}$ with respect to $\bm{m}$ and $\bm{n}$ can be found in the Appendix~\ref{AppA3}. The combination of all these identities leads to the following expression for the reactive magnetic torque in terms of the injected spin current:
\begin{widetext}
\begin{align}
\label{eq:spin_torque}
\partial_{t}m^{i}(\vec{r})&=\big\{m^{i}(\vec{r}),\mathcal{E}_{\textrm{ST}}\big\}=\int d\vec{r}\,'\left[\big\{m^{i}(\vec{r}),m^{j}(\vec{r}\,')\big\}\frac{\delta\mathcal{E}_{\textrm{ST}}}{\delta m^{j}(\vec{r}\,')}+\big\{m^{i}(\vec{r}),\Omega^{j}(\vec{r}\,')\big\}\frac{\delta\mathcal{E}_{\textrm{ST}}}{\delta n^{j}(\vec{r}\,')}\right],\\
&=-\tfrac{4m_{\parallel}g_{4}^{\textrm{AM}}}{\hbar^{2}}J_{\kappa}^{i}(\vec{r})\sigma_{x}|_{\kappa\sigma}(\bm{m}\bigcdot\partial_{\sigma}\bm{n})(\vec{r})+\tfrac{4m_{\parallel}g_{4}^{\textrm{AM}}}{\hbar^{2}}(\bm{m}\bigcdot\bm{J}_{\kappa})(\vec{r})\sigma_{x}|_{\kappa\sigma}\partial_{\sigma}n^{i}(\vec{r})+\tfrac{4m_{\parallel}g_{4}^{xy}}{\hbar^{2}}(\bm{n}\bigcdot\bm{J}_{\kappa})(\vec{r})\partial_{\kappa}n^{i}(\vec{r})\nonumber\\
&\hspace{0.5cm}-\tfrac{4m_{\parallel}g_{4}^{xy}}{\hbar^{2}}n^{l}(\vec{r})\partial_{\kappa}\big[J_{\kappa}^{i}n^{l}\big](\vec{r})+\tfrac{4m_{\parallel}g_{4}^{xy}}{\hbar^{2}}n^{l}(\vec{r})\partial_{\kappa}\big[J_{\kappa}^{l}n^{i}\big](\vec{r})-\tfrac{4m_{\parallel}g_{4}^{\textrm{AM}}}{\hbar^{2}}n^{l}(\vec{r})\sigma_{x}|_{\kappa\sigma}\partial_{\sigma}\big[J_{\kappa}^{i}m^{l}\big](\vec{r})\nonumber\\
&\hspace{0.5cm}+\tfrac{4m_{\parallel}g_{4}^{\textrm{AM}}}{\hbar^{2}}n^{l}(\vec{r})\sigma_{x}|_{\kappa\sigma}\partial_{\sigma}\big[J_{\kappa}^{l}m^{i}\big](\vec{r}),\nonumber
\end{align}
\end{widetext}
With account of the spin-splitter relation $\bm{J}_{\kappa}=\tfrac{m_{\parallel}}{m_{xy}^{s}}\tfrac{\hbar}{2e}\bm{n}\,\sigma_{x}|_{\kappa\sigma}j_{\sigma}^{e}$, see Eq.~\eqref{eq:spin_splitter2}, the above contributions to the reactive magnetic torque simplify to
\begin{equation}
\label{eq:spin_torque2}
\partial_{t}\bm{m}=\tfrac{4m_{\parallel}^{2}}{m_{xy}^{s}}\tfrac{g_{4}^{xy}\mathfrak{s}^{2}}{\hbar^{2}}\tfrac{\hbar}{2e}\big[j^{e}_{x}\partial_{y}+j^{e}_{y}\partial_{x}\big]\bm{n},
\end{equation}
where we have disregarded those terms in Eq.~\eqref{eq:spin_torque} that $i)$ depend on derivatives of the total magnetization and $ii)$ are quadratic in the spin splitting $\propto t'-t$ (via $g_{4}^{\textrm{AM}}$ and $1/m_{xy}^{s}$). We note that the dissipative (antidamping-like) contribution to the spin-transfer torque can be obtained by applying the usual phenomenological arguments; in particular, it must be transversal to the fieldlike one. Therefore, we conclude that the altermagnetic contributions to the spin-transfer torques in $d$-wave altermagnets can be cast as
\begin{equation}
\label{eq:spin_torque3}
\bm{\tau}_{\bm{m},\textrm{ST}}^{\textrm{AM}}=\eta_{\textrm{FL}}\big[j^{e}_{x}\partial_{y}+j^{e}_{y}\partial_{x}\big]\bm{n}+\tfrac{\eta_{\textrm{DL}}}{\mathfrak{s}}\bm{n}\Cross\big[j^{e}_{x}\partial_{y}+j^{e}_{y}\partial_{x}\big]\bm{n}.
\end{equation}
We remark that, contrary to the case of bipartite antiferromagnets, the fieldlike component~\eqref{eq:spin_torque2} of the spin-transfer torque is allowed by symmetry considerations: the local (nonmagnetic) environment of the magnetic atoms is microscopically distinct for each magnetic sublattice, see Fig.~\ref{Fig1}(a), which breaks the sublattice symmetry $\bm{n}\rightarrow-\bm{n}$. In our tight-binding description, this is captured by the interchange $t\leftrightarrow t'$ of hopping matrix elements between both sublattices.

The Hamilton symplectic framework is also well suited to derive the current-induced reactive torque acting upon the N\'{e}el order of a $d$-wave altermagnet. The corresponding Poisson-Liouville dynamical equation reads $\partial_{t}\bm{n}=\big\{\bm{n},\mathcal{E}_{\textrm{ST}}\big\}$ and, with account of the aforementioned Poisson bracket structure describing the localized magnetic background and the spin-splitter relation, we obtain
\begin{widetext}
\begin{align}
\label{eq:Neel_torque}
\partial_{t}n^{i}(\vec{r})&=\big\{n^{i}(\vec{r}),\mathcal{E}_{\textrm{ST}}\big\}=\int d\vec{r}\,'\left[\big\{n^{i}(\vec{r}),m^{j}(\vec{r}\,')\big\}\frac{\delta\mathcal{E}_{\textrm{ST}}}{\delta m^{j}(\vec{r}\,')}+\big\{n^{i}(\vec{r}),n^{j}(\vec{r}\,')\big\}\frac{\delta\mathcal{E}_{\textrm{ST}}}{\delta n^{j}(\vec{r}\,')}\right],\\
&=\tfrac{4m_{\parallel}g_{4}^{\textrm{AM}}}{\hbar^{2}}(\bm{n}\bigcdot\bm{J}_{\kappa})(\vec{r})\sigma_{x}\big|_{\kappa\sigma}\partial_{\sigma}n^{i}(\vec{r}),\nonumber\\
&=\tfrac{4m_{\parallel}^{2}}{m_{xy}^{s}}\tfrac{g_{4}^{\textrm{AM}}\mathfrak{s}^{2}}{\hbar^{2}}\tfrac{\hbar}{2e}\big(\vec{j}^{e}\cdot\vec{\nabla}\big
)n^{i}(\vec{r}).\nonumber
\end{align}
\end{widetext}
It is worth noting that, since this expression is qua\-dratic in the spin splitting (due to the ratio $g_{4}^{\textrm{AM}}/m_{xy}^{s}$), the effect of this spin-transfer torque on the N\'{e}el order is reduced relative to Eq.~\eqref{eq:spin_torque3} in the weak regime $t'-t\ll t_{0}+t+t'-t^{AB}$, where it will be in general disregarded. Furthermore, this same expression also exhibits the usual isotropic (in the spatial indices) coupling between the applied charge current and the differential operator $\vec{\nabla}$, akin to the bipartite antiferromagnetic case \cite{Baltz-RMP2018}.

\subsection{Spin pumping}

We discuss in this subsection the charge current generated within the bulk of a $d$-wave altermagnet by the dynamics of a spin texture present in the localized magnetic background. This pumping of spin current into the itinerant fluid is a reciprocal process (in the Onsager sense) to the induction of spin-transfer torques acting on the localized spin degrees of freedom by the flow of a charge current. We start by noting that the Landau-Lifshitz-Gilbert (LLG) system of equations for a $d$-wave altermagnet have the following general form:
\begin{align}
\label{eq:LLG}
\partial_{t}\bm{n}&=\bm{n}\Cross\bm{f}_{\bm{m}}+\bm{\tau}_{\bm{n}},\\
\partial_{t}\bm{m}+\tfrac{1}{\mathfrak{s}}\,\bm{n}\Cross\underline{\alpha}\partial_{t}\bm{n}&=\bm{m}\Cross\bm{f}_{\bm{m}}+\bm{n}\Cross\bm{f}_{\bm{n}}+\bm{\tau}_{\bm{m}},\nonumber
\end{align}
where $\bm{f}_{\bm{x}}=-\tfrac{\delta\mathcal{E}_{\textrm{d-wave}}}{\delta{\bm{x}}}$ is the thermodynamic force associated with the thermodynamic variable $\bm{x}\,(=\bm{m},\bm{n})$, and $\bm{\tau}_{\bm{x}}$ denote the magnetic torque acting upon $\bm{x}$ resultant from the interaction with other degrees of freedom (which includes, in particular, the spin-transfer torque). Here, $\mathfrak{s}$ and $\underline{\alpha}$ are the saturation spin density and the Gilbert dissipation tensor of the magnet, respectively. The latter arises from the choice $\mathcal{R}[\bm{m},\bm{n}]\equiv\frac{1}{2\mathfrak{s}}\alpha_{ij}\partial_{t}n_{i}\partial_{t}n_{j}$ for the leading contribution to the Rayleigh dissipation functional.

Thermodynamic fluxes and forces are related via the following Onsager matrix for the spin and charge sectors:
\begin{align}
\label{eq:Onsager_matrix}
\left(\begin{array}{c}
\partial_t\bm{m}\\
\vec{j}^{e}
\end{array}\right)=\left(\begin{array}{ccc}
\cdot\star\cdot & & \underline{L}^{(\textrm{sq})} \\
 \underline{L}^{(\textrm{qs})} & & \vartheta
\end{array}\right)
\left(\begin{array}{c}
\bm{f}_{\bm{m}}\\
\vec{E}
\end{array}\right),
\end{align}
where $\cdot\star\cdot$ denotes a linear-response coefficient inconsequential for our discussion. From Eq.~\eqref{eq:spin_torque3} for the altermagnetic spin-transfer torque, we can identify the matrix coefficients of the $sq$ off-diagonal block, $[\underline{L}^{(\textrm{sq})}]_{ji}=\vartheta\sigma_{x}|_{i\beta}\big[\eta_{\textrm{FL}}\partial_{\beta}n^{j}+(\eta_{\textrm{DL}}/\mathfrak{s})\epsilon_{jlk}n^{l}\partial_{\beta}n^{k}\big]$. We note here that the Onsager reciprocity principle yields the following relation between the off-diagonal blocks, $[\underline{L}^{(\textrm{qs})}]_{ij}[\bm{m},\bm{n}]=-[\underline{L}^{(\textrm{sq})}]_{ji}[-\bm{m},-\bm{n}]$. Furthermore, the thermodynamic force $\bm{f}_{\bm{m}}$ can be cast as $\bm{f}_{\bm{m}}\approx -\bm{n}\Cross\partial_{t}\bm{n}/\mathfrak{s}^{2}$ from Eq.~\eqref{eq:LLG}, where we have taken into account that its leading contribution is of the form $-\bm{m}/\chi$ and, therefore, $\bm{n}\cdot\bm{f}_{\bm{m}}=0$ at the leading order. Thus, we can now calculate the spin-pumping current induced by a dynamical magnetic texture in the background of a $d$-wave altermagnet, which is given by the term $\vec{j}^{e,\textrm{pump}}=\underline{L}^{(\textrm{qs})}\bm{f}_{\bm{m}}$ and reads
\begin{align}
\label{eq:sin-pumping_current}
j^{e,\textrm{pump}}_{x}&=\tfrac{\vartheta}{\mathfrak{s}}\big[\tfrac{\eta_{\textrm{FL}}}{\mathfrak{s}}\bm{n}\Cross\partial_{y}\bm{n}+\eta_{\textrm{DL}}\partial_{y}\bm{n}\big]\bigcdot\partial_{t}\bm{n},\\
j^{e,\textrm{pump}}_{y}&=\tfrac{\vartheta}{\mathfrak{s}}\big[\tfrac{\eta_{\textrm{FL}}}{\mathfrak{s}}\bm{n}\Cross\partial_{x}\bm{n}+\eta_{\textrm{DL}}\partial_{x}\bm{n}\big]\bigcdot\partial_{t}\bm{n}.\nonumber
\end{align}

\section{Effect of elastic deformations on the transport of spin and charge in d-wave altermagnets}

\label{Sec5}

So far, the findings discussed in the previous sections have been obtained for a pristine crystal structure, i.e. in the absence of lattice deformations. We proceed next to discuss the effect of strain on the transport properties of the itinerant fluid and, in particular, on its spin-transfer response. Within the framework of the linear-response theory, we can capture the leading contribution of strain to the transport properties of $d$-wave altermagnets by restricting ourselves to the study of the effect of lattice deformations on the (tight-binding) hopping matrix elements. To do so, we consider a simple model in which the dependence of the hopping constants $t_{ij}$ on the distance between (deformed) lattice sites is given by
\begin{align}
\label{eq:t_strain}
t_{ij}&\equiv t(|\vec{r}_{i}-\vec{r}_{j}|)= t_{ij}^{0}e^{-\left[|\vec{r}_{i}-\vec{r}_{j}|-|\vec{r}_{i}^{\,0}-\vec{r}_{j}^{\,0}|\right]/\zeta}\\
&\simeq t_{ij}^{0}-\tfrac{t_{ij}^{0}}{\zeta}\left[|\vec{r}_{i}-\vec{r}_{j}|-|\vec{r}_{i}^{\,0}-\vec{r}_{j}^{\,0}|\right],\nonumber
\end{align}
where $t_{ij}^{0}$ and $\{\vec{r}_{i}^{\,0}\}_{i}$ are the value of the hopping matrix elements and the lattice sites associated with the unstrained crystal lattice, respectively. Furthermore, $\zeta$ is a model-dependent correlation/overlap length, and we have expanded the exponential factor up to linear order in $\zeta^{-1}$. We consider next a general (homogeneous) strain applied to the crystal, parametrized by the strain tensor
\begin{equation}
\label{eq:strain_tensor}
\underline{\epsilon}=\begin{pmatrix}
\epsilon_{xx} & \epsilon_{xy} & \epsilon_{xz}\\
\epsilon_{xy} & \epsilon_{yy} & \epsilon_{yz} \\
\epsilon_{xz} & \epsilon_{yz} & \epsilon_{zz} \\
\end{pmatrix},
\end{equation}
so that the lattice sites are now located at $\vec{r}_{i}=\left[\mathds{1}+\underline{\epsilon}\right]\vec{r}_{i}^{\,0}$. Since the hopping matrix elements $t$ and $t'$ correspond to the bonds $\vec{r}_{i}^{\,0}-\vec{r}_{j}^{\,0}=\pm a(\hat{e}_{x}+\hat{e}_{y})$, and $\vec{r}_{i}^{\,0}-\vec{r}_{j}^{\,0}=\pm a(\hat{e}_{x}-\hat{e}_{y})$, respectively, we obtain the lattice deformations $t\rightarrow |\vec{r}_{i}-\vec{r}_{j}|-|\vec{r}_{i}^{\,0}-\vec{r}_{j}^{\,0}|\simeq \tfrac{a}{\sqrt{2}}(\epsilon_{xx}+\epsilon_{yy}+2\epsilon_{xy})$ and $t'\rightarrow |\vec{r}_{i}-\vec{r}_{j}|-|\vec{r}_{i}^{\,0}-\vec{r}_{j}^{\,0}|\simeq \tfrac{a}{\sqrt{2}}(\epsilon_{xx}+\epsilon_{yy}-2\epsilon_{xy})$ up to linear order in the strain coefficients. As a result, the following identity for the 'spin splitting' holds
\begin{equation}
\label{eq:AM_hopping}
t-t'=(t^{0}-t'^{0})\left[1-\tfrac{a}{\sqrt{2}\zeta}(\epsilon_{xx}+\epsilon_{yy})\right]-\sqrt{2}\tfrac{a}{\zeta}(t^{0}+t'^{0})\epsilon_{xy}.
\end{equation}
In particular, we observe that it splits into two contributions, one parametrized by the unstrained spin spiltting and the other by the average $t_{0}+t_{0}'$ of unstrained tight-binding constants. Since the altermagnetic contributions to the different physical quantities of interest (namely, gauge fields, charge/spin currents, and spin-transfer torques) are parametrized by the difference $t-t'$, we can draw the following conclusions: $i$) in the case of $t_{0}$ and $t_{0}'$ having the same sign, the difference $|t^{0}-t'^{0}|$ is small and, therefore, the presence of a shear strain in the $xy$ plane ($\epsilon_{xy}\neq0$) may enhance the altermagnetic contributions to the different physical quantities/effects discussed. On the contrary, $ii$) in the case of $t_{0}$ and $t_{0}'$ having opposite signs, the difference $|t^{0}-t'^{0}|$ dominates over the average $|t^{0}+t'^{0}|$. Thus, the effect of strain here consists of renormalizing the value of $t-t'$ only (through the 'surface strain' $\epsilon_{xx}+\epsilon_{yy}$). In particular, no significant effect on the itinerant transport of a $d$-wave altermagnet will be observed when pure shear is applied.

\begin{figure*}[ht!]
\begin{center}
\includegraphics[width=1.0\textwidth]{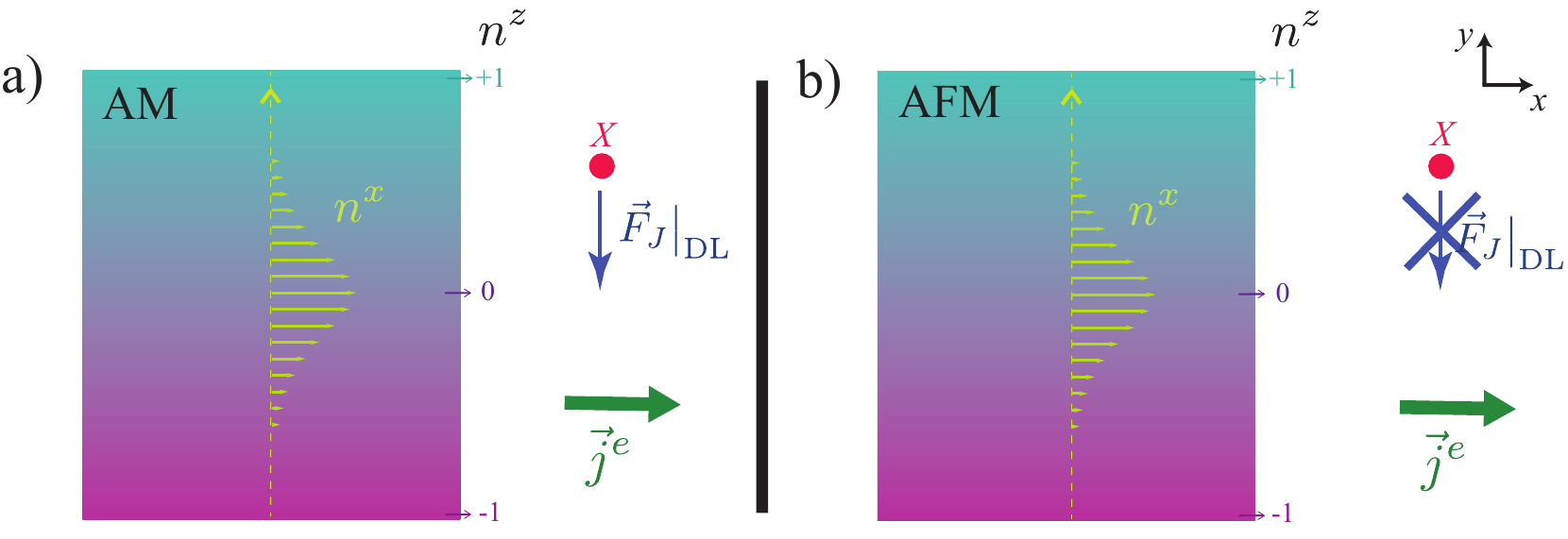}
\caption{Bloch-type domain-wall spin configuration extending along the crystallographic $y$ axis of the magnet for (a) a $d$-wave altermagnet and (b) a conventional (bipartite) antiferromagnet. Color gradient depicts the $z$-component of the (unit-norm) $s$-wave projection of the N\'{e}el order, ranging from $-1$ (fuchsia) to $+1$ (turquoise). The lime dash line illustrates the normal of the domain wall, whereas lime arrows show the $x$-component of the (unit-norm) $s$-wave projection of the N\'{e}el order. A charge current flowing along the $x$ direction of the $d$-wave altermagnet triggers the longitudinal dynamics of the domain wall (with respect to the domain-wall normal) due to the action of the dissipative component of the spin-transfer torque~\eqref{eq:spin_torque3}. In the antiferromagnetic case, however, the same current does not trigger any spin-transfer-induced dynamics on the domain wall.} 
\vspace{-0.5cm} 
\label{Fig2}
\end{center}
\end{figure*}

\section{Discussion}
\label{Discussion}

We have built an effective Yang-Mills theory for itinerant carriers in a $d$-wave altermagnet, from which we have inferred the charge and spin transport properties of the altermagnetic medium as well as elucidated the emergent spin-transfer and spin-pumping physics. We note first that our Euclidean Lagrangian~\eqref{eq:EuclidLag_AM} embodies, in the limit $t'-t\rightarrow0$, the transport theory for itinerant carriers in bipartite antiferromagnets proposed by Shraiman and Siggia in the context of high-$T_{c}$ superconductivity \cite{Shraiman-PRL1988,Shraiman-PRB1990}. Furthermore, we have observed that the spin-polarized diffusive term of our effective theory (more specifically, the off-diagonal terms of the spin-polarized mass tensor) is responsible for the transverse intertwinement of both charge and spin-currents, which yields the characteristic spin-splitter effect in $d$-wave altermagnetism. Even though our findings have been obtained from a long-wavelength expansion of the $t-J$ model around the $\bm{\Gamma}$-point, we argue that their applicability extends to other possible valleys of the $d$-wave electronic band structure. For instance, the long-wavelength expansion carried out around the $\bm{Z}$-point, modulated by the wavevector $\vec{k}_{Z}=(0,0,\tfrac{\pi}{c})$, is identical to that of Eq.~\eqref{eq:EuclidLag_AM} except for those terms depending on spatial derivatives along the $z$ direction (which pick up an extra -1 sign). Therefore, in quasi-two-dimensional ($xy$-plane) systems, the transport physics for $d$-wave altermagnets is invariant under the interchange $\bm{\Gamma}\leftrightarrow\bm{Z}$ of valleys.

Second, we have utilized in this work the mean-field picture to address electronic correlations. Our justification for this choice goes along the lines of that of Ref.~\onlinecite{Zarzuela-PRB2022}, namely the variations of the sublattice spin densities describing the localized spin background occur at the mesoscale, so that itinerant carriers interact with a uniform magnetic background at the length scale of the charge fluctuations. As a result, the slave-boson occupation numbers should be averaged at mesoscopic length scales at least. We can go beyond the mean-field treatment by assuming the smoothness of these occupation numbers over mesoscopic length scales, which, according to our microscopic derivation and in the spirit of the renormalization-group approach, translates into fluctuations leading to a spatial dependence of the coupling constants $g_{0}(\vec{r}\,)$'s, $g_{1}(\vec{r}\,)$'s, $g_{2}(\vec{r}\,)$'s, $g_{4}(\vec{r}\,)$'s, $m(\vec{r}\,)$'s and $m^{s}(\vec{r}\,)$'s of our theory.

Third, our interest in elucidating the spin-transfer response of the itinerant fluid has led to one of the main findings of this work, namely the expressions for the $d$-wave altermagnetic contributions to the spin-transfer torque and spin-pumping currents in terms of the (injected) charge current and magnetic order parameter. Remarkably, our phenomenological expressions, derived on microscopic grounds, differ from those obtained in conventional (bipartite) antiferromagnetism \cite{Baltz-RMP2018}, so that we expect unconventional current-driven dynamics of spin textures to occur in $d$-wave altermagnetic platforms: Fig.~\ref{Fig2} depicts a domain wall extending along the crystallographic $y$ axis in a $d$-wave altermagnet [Fig.~\ref{Fig2}(a)] and a conventional (bipartite) antiferromagnet [Fig.~\ref{Fig2}(b)], along with the spin-transfer torques acting on it. For a charge current flowing along the $x$ direction, the spin-transfer torque vanishes in the antiferromagnetic case, yielding no net dynamics for the soliton, whereas in the altermagnetic case there is a net spin-transfer torque acting on the domain wall, which moves along its direction. On the contrary, a charge current flowing in the $y$ direction has no effect on the altermagnetic domain wall, but triggers the motion of this soliton along the domain-wall direction in the antiferromagnetic case. 

Further insight into this exotic current-driven dynamics can be gained in the case of topological solitons by means of a collective variable approach. In the case of $d$-wave altermagnets, the Thiele equation for the center-of-mass (referred to as center hereafter) variable can be cast as \cite{Gomonay-npjSpin2024,Jin-2024}:
\begin{equation}
\label{eq:Thiele_CM}
\underline{\mathcal{M}}\ddot{\vec{X}}+\underline{\Gamma}\dot{\vec{X}}+\dot{\vec{X}}\times\vec{\mathcal{B}}=\vec{F}_{\textrm{cons}}+\vec{F}_{\textrm{ST}},
\end{equation}
where $\vec{X}$ denotes the center of the soliton considered, $\big[\underline{\mathcal{M}}\big]_{ij}=\int d\vec{r}\,\partial_{i}\bm{n}\bigcdot\partial_{j}\bm{n}$ and $\big[\underline{\Gamma}\big]_{ij}=\tfrac{\mathfrak{s}}{\chi}\int d\vec{r}\,\partial_{i}\bm{n}\underline{\alpha}\partial_{j}\bm{n}$ are the components of its effective mass and drag/dissipation tensors, respectively, and $\vec{\mathcal{B}}$ represents the topological magnetic field resulting from the effective gauge that emerges solely in altermagnets due to the presence of a nontrivial spin configuration in the background, which is responsible for the topological Hall response of the system. Furthermore, $\vec{F}_{\textrm{cons},i}=-\tfrac{\mathfrak{s}^{2}}{\chi}\int d\vec{r}\, \partial_{i}\bm{n}\bigcdot\bm{f}_{\bm{n}}$ and $F_{\textrm{ST},i}=-\tfrac{1}{\chi}\int d\vec{r}\,\big[\eta_{\textrm{DL}}\mathfrak{s}\,\partial_{i}\bm{n}\bigcdot(j_{x}\partial_{y}+j_{y}\partial_{x})\bm{n}+\eta_{\textrm{FL}}\bm{n}\bigcdot\big[\partial_{i}\bm{n}\Cross(j_{x}\partial_{y}+j_{y}\partial_{x})\bm{n}\big]\big]$ denote the spatial components of the conservative and spin-transfer forces acting on the soliton center, respectively. We note that, in the above equation, we have disregarded the coupling of the center-of-mass variable to other collective degrees of freedom for the sake of simplicity.

We observe that, again, the expression for the spin-transfer force acting on topological solitons in a $d$-wave altermagnetic medium differs from that obtained in the conventional antiferromagnetic scenario \cite{Kim-PRB2017}. In this regard, two remarks should be made: first, the fieldlike contribution to $\vec{F}_{\textrm{ST}}$, parametrized by $\eta_{\textrm{FL}}$, is forbidden by the sublattice symmetry in the case of bipartite antiferromagnets. Second, if we consider only the dynamics within the $xy$ plane, it can be cast as $F_{\textrm{ST},i}|_{\textrm{FL}}=(-1)^{i}4\pi\eta_{\textrm{FL}}\tfrac{\mathfrak{s}^{3}}{\chi}j^{e}_{i}\mathcal{Q}_{\textrm{sky}}$, where $\mathcal{Q}_{\textrm{sky}}=\tfrac{1}{4\pi\mathfrak{s}^{3}}\int d\vec{r}\,\bm{n}\bigcdot\big[\partial_{x}\bm{n}\Cross\partial_{y}\bm{n}\big]$ denotes the skyrmion charge of the localized background in the presence of a topological soliton and $(-1)^{i}=-1 (+1)$ for $i=x (y)$. Strikingly, $F_{\textrm{ST},i}|_{\textrm{FL}}$ depends isotropically on the charge current in $d$-wave altermagnets, in contrast to the cases of ferrimagnets \cite{Kim-PRB2017} and magnetoelectric antiferromagnets \cite{Zarzuela-PRB2018}, where the dependence is transverse (i.e., $F_{\textrm{ST},i}|_{\textrm{FL}}=4\pi\eta_{\textrm{FL}}\tfrac{\mathfrak{s}^{3}}{\chi}\epsilon_{ik}j^{e}_{k}\mathcal{Q}_{\textrm{sky}}$ in the latter cases). This implies, in particular, that the reactive contribution to the spin-transfer torque drives the soliton motion along the direction of the injected current, whereas its dissipative contribution triggers the soliton dynamics in the direction transverse to the current (in the absence of off-diagonal elements in the mass tensor) since $F_{\textrm{ST},x}|_{\textrm{DL}}=-\tfrac{\mathfrak{s}}{\chi}\eta_{\textrm{DL}}[\mathcal{M}_{xy}j_{x}+\mathcal{M}_{xx}j_{y}]$ and $F_{\textrm{ST},y}|_{\textrm{DL}}=-\tfrac{\mathfrak{s}}{\chi}\eta_{\textrm{DL}}[\mathcal{M}_{yy}j_{x}+\mathcal{M}_{xy}j_{y}]$.

Our effective theory and findings are focused on the exchange-dominated regime, so that relativistic contributions such as the Dzyaloshinskii-Moriya interaction have been disregarded. We note that, however, these contributions can be relevant depending on the particular site symmetry of the magnetic atoms. As a final remark, in the exchange-dominated scenario, the N\'{e}el order can exhibit an arbitrary orientation in the spin space (absence of preferential directions due to relativistic/spin-orbit effects) and the magnetic torques will follow the orientation of this order parameter.

\section{Acknowledgements}

This work has been supported by the Deutsche Forschungsgemeinschaft (DFG, German Research Foundation) - TRR 173 -- 268565370 (projects A03, A11 and B15), TRR 288 -- 422213477 (projects A12N and B05), project  445976410 (MATHEEIAS), and the Dynamics and Topology Centre TopDyn funded by the State of Rhineland Palatinate.
\newline

\textit{Note added:} During the completion of this manuscript we became aware of the work by Kokkeler \emph{et al.}\cite{Kokkeler-2024}, in which the authors develop a quantum transport theory for unconventional magnets. Their low-energy theory is based upon the generalization of the Keldysh nonlinear $\sigma$-model to these magnetic systems, whereas ours is built upon the $t-J$ model for strongly correlated systems. We have obtained similar hydrodynamic equations for the itinerant charge and spin degrees of freedom of a $d$-wave altermagnet, since these two approaches are complementary. We also became aware of the work by Vakili \emph{et al.}~\cite{Vakili-2024}, in which the authors have derived an equivalent expression for the altermagnetic contribution to the spin-transfer torque based upon phenomenological grounds.

\appendix

\section{Microscopic derivation of the long-wavelength Euclidean Lagrangian}

In this Appendix we provide a brief introduction to the slave-boson formalism, see subappendix A1, with special emphasis on its application to the tight-binding term of the $t-J$ model and the path-integral kinetic term for fermions. We also provide a detailed derivation of the low-energy long-wavelength theory of itinerant spin-charge transport for $d$-wave altermagnets in subappendix A2, which incorporates all the emergent \emph{valley} contributions. We note in passing that bold and $\vec{\,}\,$ represent vectors in the spin and real spaces, respectively, and that $\bigcdot$ and $\Cross$ denote the scalar and vector products in the spin space, respectively.

\subsection{Slave-boson formalism}
\label{AppA1}

The theory of itinerant transport developed here is built upon the spin-rotation invariant slave-boson (SRI SB) formulation of strongly correlated systems \cite{Kotliar-PRL1986,Li-PRB1989,Fresard-EPL1991,Fresard-JCM1992}. This choice is motivated by the fact that the SB formalism incorporates, in a natural and controllable way, the exchange between the itinerant and localized spin degrees of freedom into the representation of the electron operators, as it was first shown in Ref.~\onlinecite{Zarzuela-PRB2022} for the case of magnetically frustrated conductors. The starting point is the following spin-charge separation of the electron operators \cite{Kotliar-PRL1986,Li-PRB1989}
\begin{equation}
\label{eq:SRI-SB}
c_{i,\sigma}=\sum_{\sigma'}z_{i,\sigma\sigma'}f_{i,\sigma'},
\end{equation}
where $f_{i,\sigma}$ are spinless pseudofermion operators (describing the \emph{charge} degree of freedom) and the operator matrix $\underline{z_{i}}$ is defined as
\begin{widetext}
 \begin{align}
 \label{eq:matrix_z}
 \underline{z_{i}}&=[(1-d_{i}^{\dagger}d_{i})\tau_{0}-\underline{S_{i}}^{\dagger}\underline{S_{i}}]^{-1/2}\big(e_{i}^{\dagger}\underline{S_{i}}+\underline{S_{T,i}}^{\dagger}d_{i}\big)[(1-e_{i}^{\dagger}e_{i})\tau_{0}-\underline{S_{T,i}}^{\dagger}\underline{S_{T,i}}]^{-1/2}.\nonumber
 \end{align}
 \end{widetext}
Here, the spin operator $\underline{S_{i}}$ and its time-reversed counterpart $\underline{S_{T,i}}=\hat{T}\underline{S_{i}}\hat{T}^{-1}$ read 
 \begin{align}
 \underline{S_{i}}&=\tfrac{1}{\sqrt{2}}\left[s_{0,i}\tau_{0}+\bm{s}_{i}\bigcdot\bm{\tau}\right],\hspace{0.05cm}\underline{S_{T,i}}=\tfrac{1}{\sqrt{2}}\left[s_{0,i}\tau_{0}-\bm{s}_{i}\bigcdot\bm{\tau}\right],
 \end{align}
with $\tau_{0}$, $\bm{\tau}$ being the identity and the vector of Pauli matrices in the pseudospin space, respectively. The SB operators $\{e_{i},d_{i},s_{0,i},\bm{s}_{i}\}$ describe the empty, double and single occupied states at the $i$-th lattice site, respectively. It is worth noting that the constraint $e_{i}^{\dagger}e_{i}+s_{0,i}^{\dagger}s_{0,i}+\bm{s}_{i}\hspace{0.01cm}^{\dagger}\bigcdot\bm{s}_{i}+d_{i}^{\dagger}d_{i}=1$ must be imposed on the bosonic occupancy, since only one SB state is physically possible at each lattice site. This yields the identity
\begin{widetext}
 \begin{align}
 \label{eq:int_step1}
 \underline{S_{i}}^{\dagger}\underline{S_{i}}&=\tfrac{1}{2}\big[(1-d_{i}^{\dagger}d_{i}-e_{i}^{\dagger}e_{i})\tau_{0}+(s_{0,i}^{\dagger}\bm{s}_{i}+\bm{s}_{i}\hspace{0.01cm}^{\dagger}s_{0,i}+i\bm{s}_{i}\hspace{0.01cm}^{\dagger}\hspace{-0.05cm}\Cross\bm{s}_{i})\bigcdot\bm{\tau}\big]
 \end{align}
 \end{widetext}
 for the (self-adjoint) product of spin operators. As a result, the tight-binding term of the $t-J$ model in the SRI SB representation becomes
\begin{align}
\label{eq:tb_hamiltonian}
H_{\textrm{tb}}^{\textrm{SRI SB}}&=\sum_{\langle i,j\rangle,\sigma_{1},\sigma_{2}}t_{ij}(f_{i,\sigma_{1}}^{\dagger}z_{i,\sigma_{1}\sigma}^{\dagger})(z_{j,\sigma\sigma_{2}}f_{j,\sigma_{2}})\\
&=\frac{1}{2}\sum_{i, j\in\textrm{NN}(i)}t_{ij}\Psi_{i}^{\dagger}\big(\underline{z_{i}}^{\dagger}\underline{z_{j}}\big)\Psi_{j}+\textrm{h.c.},\nonumber
\end{align}
where, again, $\Psi_{i}$ denotes the spinor describing the itinerant carrier at the $i$-th site (namely, $\Psi_{i,\sigma}\equiv f_{i,\sigma}$). In the spirit of Refs.~\onlinecite{Li-PRB1989}, \onlinecite{Fresard-EPL1991} and~\onlinecite{Fresard-JCM1992}, we assume that the spin densities $\{\bm{S}^{\alpha}\}_{\alpha=A,B}$ describing the magnetic sublattices $\{\Lambda_{\alpha}\}_{\alpha=A,B}$ of the localized background dictate the spatial dependence of the slave Bose fields: those Bose fields transforming as scalars in the spin space will be taken  as spatially homogeneous, namely $\langle e_{i}^{\dagger}\rangle_{\textrm{MF}}=\langle e_{i}\rangle_{\textrm{MF}}\equiv e$, $\langle d_{i}^{\dagger}\rangle_{\textrm{MF}}=\langle d_{i}\rangle_{\textrm{MF}}\equiv d$, and $\langle s_{0,i}^{\dagger}\rangle_{\textrm{MF}}=\langle s_{0,i}\rangle_{\textrm{MF}}\equiv s_{0}$. However, those Bose fields transforming as vectors will adjust to the localized spin background adiabatically, namely $\langle \bm{s}_{i}\hspace{0.02cm}^{\dagger}\rangle_{\textrm{MF}}=\langle \bm{s}_{i}\rangle_{\textrm{MF}}\equiv\zeta\bm{S}^{\alpha}(\vec{r}_{i})$ if $\vec{r}_{i}\in\Lambda_{\alpha}$. Here, $\zeta$ is a mean-field parameter describing the length of the bosonic spin triplet. Therefore, the following expression for the mean-field value of the operator matrix product $\underline{z_{i}}^{\dagger}\underline{z_{j}}$ is obtained \cite{Zarzuela-PRB2022}:
\begin{widetext}
 \begin{align}
 \label{eq:matrix_z_prod}
 \langle\underline{z_{i}}^{\dagger}\underline{z_{j}}\rangle_{\textrm{MF}}&=\Big[A_{1}^{2}+A_{2}^{2}\,\bm{S}^{\alpha}(\vec{r}_{i})\bigcdot\bm{S}^{\beta}(\vec{r}_{j})\Big]\tau_{0}+\Big[A_{1}A_{2}\big(\bm{S}^{\alpha}\big(\vec{r}_{i})+\bm{S}^{\beta}(\vec{r}_{j})\big)+iA_{2}^{2}\,\bm{S}^{\alpha}(\vec{r}_{i})\Cross\bm{S}^{\beta}(\vec{r}_{j})\Big]\bigcdot\bm{\tau},
 \end{align}
 \end{widetext}
where we have considered that $\vec{r}_{i}\in\Lambda_{\alpha}$ and $\vec{r}_{j}\in\Lambda_{\beta}$. The above coefficients $A_{1}$ and $A_{2}$ take the form \cite{Fresard-EPL1991,Fresard-JCM1992}
\begin{widetext}
\begin{align}
\label{eq:A1}
A_{1}(s_{0},\mathfrak{s},e,d)&\equiv\frac{1}{\sqrt{2}}\Big(s_{0}(e+d)\big[a_{(+,-)}a_{(-,+)}+a_{(+,+)}a_{(-,-)}\big]+\zeta\,\mathfrak{s}(e-d)\big[a_{(+,-)}a_{(-,+)}-a_{(+,+)}a_{(-,-)}\big]\Big),\\
\label{eq:A2}
 A_{2}(s_{0},\mathfrak{s},e,d)&\equiv\frac{1}{\sqrt{2}}\Big(\frac{s_{0}}{\mathfrak{s}}(e+d)\big[a_{(+,-)}a_{(-,+)}-a_{(+,+)}a_{(-,-)}\big]+\zeta(e-d)\big[a_{(+,-)}a_{(-,+)}+a_{(+,+)}a_{(-,-)}\big]\Big),
 \end{align}
\end{widetext}
in terms of the auxiliary function $a_{(\sigma_{1},\sigma_{2})}\equiv\big[1+\sigma_{1}(e^{2}-d^{2})+2\sigma_{2}s_{0}\zeta\mathfrak{s}\big]^{-1/2}$. It is worth remarking here that the norm $|\bm{S}^{\alpha}(\vec{r})|$ is not uniform across the system due to the multipolar behavior exhibited by the sublattice spin densities. In this regard, we assume that, in the spirit of the long-wavelength expansion of the tight-binding term, the spatial inhomogeneities of the norm $|\bm{S}^{\alpha}(\vec{r})|$ contribute at the subleading order, namely we approximate $|\bm{S}^{\alpha}(\vec{r})|\approx\tfrac{1}{V}\int_{V}d\vec{r}\,|\bm{S}^{\alpha}(\vec{r})|\equiv\mathfrak{s}$ by its volume average. As a result, Eq.~\eqref{eq:tb_hamiltonian} becomes
\begin{widetext}
\begin{align}
\label{eq:tb_hamiltonian2}
H_{\textrm{tb}}^{\textrm{SRI SB}}&=\frac{1}{2}\sum_{i,j}t_{ij}\Big[A_{1}^{2}+A_{2}^{2}\,\bm{S}^{\alpha}(\vec{r}_{i})\bigcdot\bm{S}^{\beta}(\vec{r}_{j})\Big]\Psi_{i}^{\dagger}\Psi_{j}+
\frac{1}{2}\sum_{i,j}t_{ij}\Big[A_{1}A_{2}\big(\bm{S}^{\alpha}\big(\vec{r}_{i})+\bm{S}^{\beta}(\vec{r}_{j})\big)+iA_{2}^{2}\,\bm{S}^{\alpha}(\vec{r}_{i})\Cross\bm{S}^{\beta}(\vec{r}_{j})\Big]\bigcdot\Psi_{i}^{\dagger}\bm{\tau}\Psi_{j}.
\end{align}
\end{widetext}

Similarly, we can apply the SRI SB treatment to the kinetic Lagrangian of a fermionic system, rooted in the microscopic path-integral term $\frac{\hbar}{2}\sum_{i,\sigma}\big[\hat{c}_{i,\sigma}^{\dagger}\partial_{0}\hat{c}_{i,\sigma}-\partial_{0}\hat{c}_{i,\sigma}^{\dagger}c_{i,\sigma}\big]$, where $x_{0}=it$ denotes the Wick-rotated time coordinate. Its mean-field expression takes the form
\begin{widetext}
\begin{align}
\label{eq:kin_term_init}
\Big\langle\sum_{i,\sigma}\big(c_{i,\sigma}^{\dagger}\partial_{0}c_{i,\sigma}-\partial_{0}c_{i,\sigma}^{\dagger}c_{i,\sigma})\Big\rangle_{\textrm{MF}}&=\sum_{i}\Psi_{i}^{\dagger}\big[\langle\underline{z_{i}}\rangle_{\textrm{MF}}^{\dagger}\partial_{0}\langle\underline{z_{i}}\rangle_{\textrm{MF}}-\partial_{0}\langle\underline{z_{i}}\rangle_{\textrm{MF}}^{\dagger}\langle\underline{z_{i}}\rangle_{\textrm{MF}}\big]\Psi_{i}\\
&\hspace{1cm}+\sum_{i}\Big[\Psi_{i}^{\dagger}\langle\underline{z_{i}}^{\dagger}\underline{z_{i}}\rangle_{\textrm{MF}}\partial_{0}\Psi_{i}-\partial_{0}\Psi_{i}^{\dagger}\langle\underline{z_{i}}^{\dagger}\underline{z_{i}}\rangle_{\textrm{MF}}\Psi_{i}\Big].\nonumber
\end{align}
\end{widetext}

With account of the identities $\langle\underline{z_{i}}^{\dagger}\underline{z_{i}}\rangle_{\textrm{MF}}=\big(A_{1}^{2}+A_{2}^{2}\mathfrak{s}^{2}\big)\tau_{0}+2A_{1}A_{2}\bm{S}^{\alpha}(\vec{r}_{i})\bigcdot\bm{\tau}$, $\langle\underline{z_{i}}\rangle_{\textrm{MF}}^{\dagger}\partial_{0}\langle\underline{z_{i}}\rangle_{\textrm{MF}}=\Big[A_{1}A_{2}\partial_{0}\bm{S}^{\alpha}+iA_{2}^{2}\bm{S}^{\alpha}\Cross\partial_{0}\bm{S}^{\alpha}\Big](\vec{r}_{i})\bigcdot\bm{\tau}$, and $\partial_{0}\langle\underline{z_{i}}\rangle_{\textrm{MF}}^{\dagger}\langle\underline{z_{i}}\rangle_{\textrm{MF}}=\Big[A_{1}A_{2}\partial_{0}\bm{S}^{\alpha}-iA_{2}^{2}\bm{S}^{\alpha}\Cross\partial_{0}\bm{S}^{\alpha}\Big](\vec{r}_{i})\bigcdot\bm{\tau}$, in which we have assumed that $\vec{r}_{i}\in\Lambda_{\alpha}$, we obtain the following expression for the mean-field path-integral kinetic term:
\begin{widetext}
\begin{align}
\label{eq:kin_term}
\Big\langle\tfrac{\hbar}{2}\sum_{i,\sigma}\big(c_{i,\sigma}^{\dagger}\partial_{0}c_{i,\sigma}-\partial_{0}c_{i,\sigma}^{\dagger}c_{i,\sigma}\big)\Big\rangle_{\textrm{MF}}&=\sum_{i}\Big\{\tfrac{\hbar}{2}(A_{1}^{2}+A_{2}^{2}\mathfrak{s}^{2})\big[\Psi^{\dagger}\partial_{0}\Psi-\partial_{0}\Psi^{\dagger}\Psi\big](\vec{r}_{i})+2A_{1}A_{2}\,\bm{S}^{\alpha}(\vec{r}_{i})\bigcdot\big[\Psi^{\dagger}\bm{\tau}\partial_{0}\Psi-\partial_{0}\Psi^{\dagger}\bm{\tau}\Psi\big](\vec{r}_{i})\nonumber\\
&\hspace{1cm}+2iA_{2}^{2}[\bm{S}^{\alpha}\Cross\partial_{0}\bm{S}^{\alpha}](\vec{r}_{i})\bigcdot\big[\Psi^{\dagger}\tfrac{\hbar}{2}\bm{\tau}\Psi\big](\vec{r}_{i})\Big\}.
\end{align}
\end{widetext}

\subsection{Microscopic derivation of the long-wavelength Lagrangian for the itinerant carriers}

\label{AppA2}

In this Section we formally derive the effective Euclidean Lagrangian~\eqref{eq:EuclidLag_AM} describing the dynamics of itinerant carriers in a $d$-wave altermagnet from the single-band doped tight-binding term of the $t-J$ model via the SRI SB representation of the electron operators. With RuO$_{2}$ in mind, we consider the microscopic Hamiltonian to be defined on a rutile crystal lattice, see Fig.~\ref{Fig1}(a) for details on the lattice parameters, bond vectors and relevant hopping matrix elements. We note that, in our model, the anisotropy in the tight-binding constants $t\neq t'$ encapsulates the $d$-wave altermagnetic character of the system \cite{Smejkal-PRL2023}. As depicted in Fig.~\ref{Fig1}(a), we can describe the magnetic ground state of a $d$-wave altermagnet via a single unit cell (contoured in grey), consisting of two 'atoms' belonging to the two magnetic sublattices $\Lambda_{\textrm{A}},\Lambda_{\textrm{B}}$, which repeats itself along the sites $\vec{r}_{k}\in\Lambda_{\textrm{m}}$ of the corresponding (magnetic) crystal lattice. Remarkably, the spin densities centered at the 'atoms' in the same unit cell 1) have opposite sign and are connected only via crystal rotations (e.g., $C_{4,(001)}$ in the case of RuO$_{2}$) combined with a translation, and 2) their projection onto a quantization axis in the spin space yields \emph{anisotropic} isosurfaces in the direct space, see Fig.~\ref{Fig1}(b). Within a given sublattice, the projection axis of these spin-density isosurfaces may change (smoothly) between neighboring unit cells in the long-wavelength limit. In what follows, we will describe the localized spin sector in the continuum regime by means of two smooth fields, namely the macroscopic spin density $\bm{m}(\vec{r})\equiv\tfrac{1}{2}\big[\bm{S}^{A}+\bm{S}^{B}\big](\vec{r})$ and the N\'{e}el order $\bm{n}(\vec{r})\equiv\tfrac{1}{2}\big[\bm{S}^{A}-\bm{S}^{B}\big](\vec{r})$. Similarly, the itinerant fluid will be described, at each valley, by a spinor splitting into the product of a planar wave and a smooth envelope, $\Psi(\vec{r})\equiv e^{i\vec{k}_{\nu}\cdot\vec{r}}\Psi_{\nu}(\vec{r})$. The planar wave function is included into our description to properly account for the charge oscillations occurring at short length scales, of the order of $1/|\vec{k}_{\nu}|$, where $\vec{k}_{\nu}$ denotes the wavevector parametrizing the $\nu$-th valley.

We can embark on the derivation of the effective Lagrangian by expanding the sublattice spin densities $\bm{S}^{A,B}(\vec{r}_{j})$ and the Fermi field $\Psi(\vec{r}_{j})$ around the lattice position $\vec{r}_{i}$ up to second order in the lattice parameter, therefore obtaining the expressions:
\begin{widetext}
\begin{align}
&\bm{S}^{\beta}(\vec{r}_{j})=\bm{S}^{\beta}(\vec{r}_{i})+\partial_{k}\bm{S}^{\beta}(\vec{r}_{i})[\vec{e}_{ij}]_{k}+\tfrac{1}{2}\partial_{k_{1}}\partial_{k_{2}}\bm{S}^{\beta}(\vec{r}_{i})[\vec{e}_{ij}]_{k_{1}}[\vec{e}_{ij}]_{k_{2}}+\ldots,\\
&e^{i\vec{k}_{\nu}\cdot\vec{r}_{j}}\Psi_{\nu}(\vec{r}_{j})=e^{i\vec{k}_{\nu}\cdot\vec{r}_{i}}e^{i\vec{k}_{\nu}\cdot\vec{e}_{ij}}\Big[\Psi_{\nu}(\vec{r}_{i})+\partial_{k}\Psi_{\nu}(\vec{r}_{i})[\vec{e}_{ij}]_{k}+\tfrac{1}{2}\partial_{k_{1}}\partial_{k_{2}}\Psi_{\nu}(\vec{r}_{i})[\vec{e}_{ij}]_{k_{1}}[\vec{e}_{ij}]_{k_{2}}+\ldots\Big],\nonumber
\end{align}
\end{widetext}
where $\vec{e}_{ij}\equiv\vec{r}_{j}-\vec{r}_{i}$ denote the bond vectors between lattice sites. By incorporating these expansions into Eq.~\eqref{eq:tb_hamiltonian2}, the mean-field SRI SB expression for the tight-binding Hamiltonian splits into the following two contributions:
\begin{widetext}
\begin{align}
\label{eq:tb_hamiltonian_t1}
H_{\textrm{tb},\nu}^{1}&=\tfrac{1}{2}\sum_{i\in\Lambda_{\alpha}}\sum_{t^{\alpha\beta}}\sum_{\vec{e}_{ij}\in\mathcal{N}(i,t^{\alpha\beta})}\hspace{-0.4cm}t^{\alpha\beta}\Big\{e^{i\vec{k}_{\nu}\cdot\vec{e}_{ij}}\big[A_{1}^{2}+A_{2}^{2}\bm{S}^{\alpha}\bigcdot\bm{S}^{\beta}\big](\vec{r}_{i})\big[\Psi^{\dagger}_{\nu}\Psi_{\nu}\big](\vec{r}_{i})+e^{i\vec{k}_{\nu}\cdot\vec{e}_{ij}}\big[A_{1}^{2}+A_{2}^{2}\bm{S}^{\alpha}\bigcdot\bm{S}^{\beta}\big](\vec{r}_{i})\big[\Psi^{\dagger}_{\nu}\partial_{k}\Psi_{\nu}\big](\vec{r}_{i})[\vec{e}_{ij}]_{k}\nonumber\\
&\hspace{1cm}+\tfrac{1}{2}e^{i\vec{k}_{\nu}\cdot\vec{e}_{ij}}\big[A_{1}^{2}+A_{2}^{2}\bm{S}^{\alpha}\bigcdot\bm{S}^{\beta}\big](\vec{r}_{i})\big[\Psi^{\dagger}_{\nu}\partial_{k_{1}}\partial_{k_{2}}\Psi_{\nu}\big](\vec{r}_{i})[\vec{e}_{ij}]_{k_{1}}[\vec{e}_{ij}]_{k_{2}}+A_{2}^{2}e^{i\vec{k}_{\nu}\cdot\vec{e}_{ij}}\bm{S}^{\alpha}\bigcdot\partial_{k}\bm{S}^{\beta}(\vec{r}_{i})\big[\Psi^{\dagger}_{\nu}\Psi_{\nu}\big](\vec{r}_{i})[\vec{e}_{ij}]_{k}\nonumber\\
&\hspace{1cm}+A_{2}^{2}e^{i\vec{k}_{\nu}\cdot\vec{e}_{ij}}\bm{S}^{\alpha}\bigcdot\partial_{k_{1}}\bm{S}^{\beta}(\vec{r}_{i})\big[\Psi^{\dagger}_{\nu}\partial_{k_{2}}\Psi_{\nu}\big](\vec{r}_{i})[\vec{e}_{ij}]_{k_{1}}[\vec{e}_{ij}]_{k_{2}}+\tfrac{1}{2}A_{2}^{2}e^{i\vec{k}_{\nu}\cdot\vec{e}_{ij}}\bm{S}^{\alpha}\bigcdot\partial_{k_{1}}\partial_{k_{2}}\bm{S}^{\beta}(\vec{r}_{i})\big[\Psi^{\dagger}_{\nu}\Psi_{\nu}\big](\vec{r}_{i})[\vec{e}_{ij}]_{k_{1}}[\vec{e}_{ij}]_{k_{2}}\nonumber\\
&\hspace{1cm}+O(\partial^{3})\Big\}+\textrm{h.c.}
\end{align}

\begin{align}
\label{eq:tb_hamiltonian_t2}
H_{\textrm{tb},\nu}^{2}&=\tfrac{1}{2}\sum_{i\in\Lambda_{\alpha}}\sum_{t^{\alpha\beta}}\sum_{\vec{e}_{ij}\in\mathcal{N}(i,t^{\alpha\beta})}\hspace{-0.4cm}t^{\alpha\beta}\Big\{A_{1}A_{2}e^{i\vec{k}_{\nu}\cdot\vec{e}_{ij}}\big[\bm{S}^{\alpha}+\bm{S}^{\beta}\big](\vec{r}_{i})\bigcdot\big[\Psi^{\dagger}_{\nu}\bm{\tau}\Psi_{\nu}\big](\vec{r}_{i})+A_{1}A_{2}e^{i\vec{k}_{\nu}\cdot\vec{e}_{ij}}\big[\bm{S}^{\alpha}+\bm{S}^{\beta}\big](\vec{r}_{i})\bigcdot\big[\Psi^{\dagger}_{\nu}\bm{\tau}\partial_{k}\Psi_{\nu}\big](\vec{r}_{i})[\vec{e}_{ij}]_{k}\nonumber\\
&\hspace{1cm}+\tfrac{1}{2}A_{1}A_{2}e^{i\vec{k}_{\nu}\cdot\vec{e}_{ij}}\big[\bm{S}^{\alpha}+\bm{S}^{\beta}\big](\vec{r}_{i})\bigcdot\big[\Psi^{\dagger}_{\nu}\partial_{k_{1}}\partial_{k_{2}}\Psi_{\nu}\big](\vec{r}_{i})[\vec{e}_{ij}]_{k_{1}}[\vec{e}_{ij}]_{k_{2}}+A_{1}A_{2}e^{i\vec{k}_{\nu}\cdot\vec{e}_{ij}}\partial_{k}\bm{S}^{\beta}(\vec{r}_{i})\bigcdot\big[\Psi^{\dagger}_{\nu}\bm{\tau}\Psi_{\nu}\big](\vec{r}_{i})[\vec{e}_{ij}]_{k}\nonumber\\
&\hspace{1cm}+A_{1}A_{2}e^{i\vec{k}_{\nu}\cdot\vec{e}_{ij}}\partial_{k_{1}}\bm{S}^{\beta}(\vec{r}_{i})\bigcdot\big[\Psi^{\dagger}_{\nu}\bm{\tau}\partial_{k_{2}}\Psi_{\nu}\big](\vec{r}_{i})[\vec{e}_{ij}]_{k_{1}}[\vec{e}_{ij}]_{k_{2}}+\tfrac{1}{2}A_{1}A_{2}e^{i\vec{k}_{\nu}\cdot\vec{e}_{ij}}\partial_{k_{1}}\partial_{k_{2}}\bm{S}^{\beta}(\vec{r}_{i})\bigcdot\big[\Psi^{\dagger}_{\nu}\bm{\tau}\Psi_{\nu}\big](\vec{r}_{i})[\vec{e}_{ij}]_{k_{1}}[\vec{e}_{ij}]_{k_{2}}\nonumber\\
&\hspace{1cm}+iA_{2}^{2}e^{i\vec{k}_{\nu}\cdot\vec{e}_{ij}}\big[\bm{S}^{\alpha}\Cross\bm{S}^{\beta}\big](\vec{r}_{i})\bigcdot\big[\Psi^{\dagger}_{\nu}\bm{\tau}\Psi_{\nu}\big](\vec{r}_{i})+iA_{2}^{2}e^{i\vec{k}_{\nu}\cdot\vec{e}_{ij}}\big[\bm{S}^{\alpha}\Cross\bm{S}^{\beta}\big](\vec{r}_{i})\bigcdot\big[\Psi^{\dagger}_{\nu}\bm{\tau}\partial_{k}\Psi_{\nu}\big](\vec{r}_{i})[\vec{e}_{ij}]_{k}\nonumber\\
&\hspace{1cm}+\tfrac{i}{2}A_{2}^{2}e^{i\vec{k}_{\nu}\cdot\vec{e}_{ij}}\big[\bm{S}^{\alpha}\Cross\bm{S}^{\beta}\big](\vec{r}_{i})\bigcdot\big[\Psi^{\dagger}_{\nu}\bm{\tau}\partial_{k_{1}}\partial_{k_{2}}\Psi_{\nu}\big](\vec{r}_{i})[\vec{e}_{ij}]_{k_{1}}[\vec{e}_{ij}]_{k_{2}}+iA_{2}^{2}e^{i\vec{k}_{\nu}\cdot\vec{e}_{ij}}\big[\bm{S}^{\alpha}\Cross\partial_{k}\bm{S}^{\beta}\big](\vec{r}_{i})\bigcdot\big[\Psi^{\dagger}_{\nu}\bm{\tau}\Psi_{\nu}\big](\vec{r}_{i})[\vec{e}_{ij}]_{k}\nonumber\\
&\hspace{1cm}+iA_{2}^{2}e^{i\vec{k}_{\nu}\cdot\vec{e}_{ij}}\big[\bm{S}^{\alpha}\Cross\partial_{k_{1}}\bm{S}^{\beta}\big](\vec{r}_{i})\bigcdot\big[\Psi^{\dagger}_{\nu}\bm{\tau}\partial_{k_{2}}\Psi_{\nu}\big](\vec{r}_{i})[\vec{e}_{ij}]_{k_{1}}[\vec{e}_{ij}]_{k_{2}}\nonumber\\
&\hspace{1cm}+\tfrac{i}{2}A_{2}^{2}e^{i\vec{k}_{\nu}\cdot\vec{e}_{ij}}\big[\bm{S}^{\alpha}\Cross\partial_{k_{1}}\partial_{k_{2}}\bm{S}^{\beta}\big](\vec{r}_{i})\bigcdot\big[\Psi^{\dagger}_{\nu}\bm{\tau}\Psi_{\nu}\big](\vec{r}_{i})[\vec{e}_{ij}]_{k_{1}}[\vec{e}_{ij}]_{k_{2}}+O(\partial^{3})\Big\}+\textrm{h.c.}
\end{align}
\end{widetext}

We specify next the above expressions for the set of hopping matrix elements $\{t_{\alpha\beta}\}$ associated with the rutile crystal structure depicted in Fig.~\eqref{Fig1}(a). More specifically, we consider the tight-binding constants $t_{z}$, $t_{0}$, $t$ and $t'$ parametrizing intralattice--$A$ electron hoppings over the lattice bonds $\{\pm c\hat{e}_{z}\}$, $\{\pm a\hat{e}_{x}, \pm a\hat{e}_{y}\}$, $\{\pm a(\hat{e}_{x}+\hat{e}_{y})\}$ and $\{\pm a(\hat{e}_{x}-\hat{e}_{y})\}$, respectively. Similarly, intralattice--$B$ electron motion is parametrized by the same set of hopping matrix elements over the same lattice bonds, with the caveat that $t$ and $t'$ must be interchanged. We note that, in our model, the altermagnetic character lies in the difference between the hopping matrix elements $t\neq t'$ and their interchange between magnetic sublattices. We also include interlattice electron hoppings, whose leading contribution is parametrized by the tight-binding constants $t^{AB}$ over the eight lattice bonds $\big\{(\pm \tfrac{a}{2},\pm \tfrac{a}{2},\pm\tfrac{c}{2})\big\}$. The continuum limit of our tight-binding Hamiltonian in the SRI SB representation is derived by applying the Riemann's prescription $\sum_{i\in\Lambda_{A(B)}}\simeq\frac{1}{\mathscr{v}_{\textrm{uc}}}\int d\vec{r}$ to Eqs.~\eqref{eq:tb_hamiltonian_t1} and~\eqref{eq:tb_hamiltonian_t2}, with $\mathscr{v}_{\textrm{uc}}$ being the volume of the rutile unit cell. As a result, we obtain the following long-wavelength expansions for the intralattice contributions:
\begin{widetext}
\begin{align}
\label{eq:tb_expansion1_tz}
\mathcal{E}_{\textrm{tb},\nu}^{1,\textrm{intra}}\big|_{t_{z}}&=\frac{t_{z}}{\mathscr{v}_{\textrm{uc}}}\int d\vec{r}\,\Big\{\big[A_{1}^{2}+A_{2}^{2}(\bm{m}^{2}+\bm{n}^{2})\big]\big[2\cos(k_{\nu,z}c)\Psi_{\nu}^{\dagger}\Psi_{\nu}+2c\sin(k_{\nu,z}c)i\Psi_{\nu}^{\dagger}\partial_{z}\Psi_{\nu}+c^{2}\cos(k_{\nu,z}c)\Psi_{\nu}^{\dagger}\partial_{z}^{2}\Psi_{\nu}\big]\\
&\hspace{0.5cm}+A_{2}^{2}\partial_{z}\big[\bm{m}^{2}+\bm{n}^{2}\big]c^{2}\cos(k_{\nu,z}c)\Psi_{\nu}^{\dagger}\partial_{z}\Psi_{\nu}+A_{2}^{2}\big[\bm{m}\bigcdot\partial_{z}^{2}\bm{m}+\bm{n}\bigcdot\partial_{z}^{2}\bm{n}\big]c^{2}\cos(k_{\nu,z}c)\Psi_{\nu}^{\dagger}\Psi_{\nu}\Big\}+\textrm{h.c.}\nonumber
\end{align}

\begin{align}
\label{eq:tb_expansion2_tz}
\mathcal{E}_{\textrm{tb},\nu}^{2,\textrm{intra}}\big|_{t_{z}}&=\frac{t_{z}}{\mathscr{v}_{\textrm{uc}}}\int d\vec{r}\,\Big\{2A_{1}A_{2}\bm{m}\bigcdot\big[2\cos(k_{\nu,z}c)\Psi_{\nu}^{\dagger}\bm{\tau}\Psi_{\nu}+2c\sin(k_{\nu,z}c)i\Psi_{\nu}^{\dagger}\bm{\tau}\partial_{z}\Psi_{\nu}+c^{2}\cos(k_{\nu,z}c)\Psi_{\nu}^{\dagger}\bm{\tau}\partial_{z}^{2}\Psi_{\nu}\big]\\
&\hspace{-0.5cm}+2A_{1}A_{2}c^{2}\cos(k_{\nu,z}c)\partial_{z}\bm{m}\bigcdot \big(\Psi_{\nu}^{\dagger}\bm{\tau}\partial_{z}\Psi_{\nu}\big)+A_{1}A_{2}c^{2}\cos(k_{\nu,z}c)\partial_{z}^{2}\bm{m}\bigcdot(\Psi_{\nu}^{\dagger}\bm{\tau}\Psi_{\nu})\nonumber\\
&\hspace{-0.5cm}+2A_{2}^{2}\big[\bm{m}\Cross\partial_{z}\bm{m}+\bm{n}\Cross\partial_{z}\bm{n}\big]\bigcdot\big[-c\sin(k_{\nu,z}c)\Psi_{\nu}^{\dagger}\bm{\tau}\Psi_{\nu}+c^{2}\cos(k_{\nu,z}c)i\Psi_{\nu}^{\dagger}\bm{\tau}\partial_{z}\Psi_{\nu}\big]+\textrm{h.c.}\nonumber
\end{align}

\begin{align}
\label{eq:tb_expansion1_t0}
\mathcal{E}_{\textrm{tb},\nu}^{1,\textrm{intra}}\big|_{t_{0}}&=\frac{t_{0}}{\mathscr{v}_{\textrm{uc}}}\int d\vec{r}\,\Big\{\big[A_{1}^{2}+A_{2}^{2}(\bm{m}^{2}+\bm{n}^{2})\big]\big[2(\cos(k_{\nu,x}a)+\cos(k_{\nu,x}a))\Psi_{\nu}^{\dagger}\Psi_{\nu}+2a\sin(k_{\nu,x}a)i\Psi_{\nu}^{\dagger}\partial_{x}\Psi_{\nu}\\
&\hspace{-0.5cm}+2a\sin(k_{\nu,y}a)i\Psi_{\nu}^{\dagger}\partial_{y}\Psi_{\nu}+a^{2}\cos(k_{\nu,x}a)\Psi_{\nu}^{\dagger}\partial_{x}^{2}\Psi_{\nu}+a^{2}\cos(k_{\nu,y}a)\Psi_{\nu}^{\dagger}\partial_{y}^{2}\Psi_{\nu}\big]\nonumber\\
&\hspace{-0.5cm}+A_{2}^{2}\partial_{x}\big[\bm{m}^{2}+\bm{n}^{2}\big]a^{2}\cos(k_{\nu,x}a)\Psi_{\nu}^{\dagger}\partial_{x}\Psi_{\nu}+A_{2}^{2}\partial_{y}\big[\bm{m}^{2}+\bm{n}^{2}\big]a^{2}\cos(k_{\nu,y}a)\Psi_{\nu}^{\dagger}\partial_{y}\Psi_{\nu}\nonumber\\
&\hspace{-0.5cm}+A_{2}^{2}a^{2}\big[\cos(k_{\nu,x}a)\big[\bm{m}\bigcdot\partial_{x}^{2}\bm{m}+\bm{n}\bigcdot\partial_{x}^{2}\bm{n}\big]+\cos(k_{\nu,y}a)\big[\bm{m}\bigcdot\partial_{y}^{2}\bm{m}+\bm{n}\bigcdot\partial_{y}^{2}\bm{n}\big]\big)\Psi_{\nu}^{\dagger}\Psi_{\nu}\Big\}+\textrm{h.c.},\nonumber
\end{align}

\begin{align}
\label{eq:tb_expansion2_t0}
\mathcal{E}_{\textrm{tb},\nu}^{2,\textrm{intra}}\big|_{t_{0}}&=\frac{t_{0}}{\mathscr{v}_{\textrm{uc}}}\int d\vec{r}\,\Big\{2A_{1}A_{2}\bm{m}\bigcdot\big[2(\cos(k_{\nu,x}a)+\cos(k_{\nu,y}a))\Psi_{\nu}^{\dagger}\bm{\tau}\Psi_{\nu}+2a\sin(k_{\nu,x}a)i\Psi_{\nu}^{\dagger}\bm{\tau}\partial_{x}\Psi_{\nu}\\
&\hspace{-0.5cm}+2a\sin(k_{\nu,y}a)i\Psi_{\nu}^{\dagger}\bm{\tau}\partial_{y}\Psi_{\nu}+a^{2}\cos(k_{\nu,x}a)\Psi_{\nu}^{\dagger}\bm{\tau}\partial_{x}^{2}\Psi_{\nu}+a^{2}\cos(k_{\nu,y}a)\Psi_{\nu}^{\dagger}\bm{\tau}\partial_{y}^{2}\Psi_{\nu}\big]\nonumber\\
&\hspace{-0.5cm}+2A_{1}A_{2}a^{2}\cos(k_{\nu,x}a)\partial_{x}\bm{m}\bigcdot\big(\Psi_{\nu}^{\dagger}\bm{\tau}\partial_{x}\Psi_{\nu}\big)+2A_{1}A_{2}a^{2}\cos(k_{\nu,y}a)\partial_{y}\bm{m}\bigcdot\big(\Psi_{\nu}^{\dagger}\bm{\tau}\partial_{y}\Psi_{\nu}\big)\nonumber\\
&\hspace{-0.5cm}+A_{1}A_{2}a^{2}\big[\cos(k_{\nu,x}a)\partial_{x}^{2}\bm{m}+\cos(k_{\nu,y}a)\partial_{y}^{2}\bm{m}\big]\bigcdot(\Psi_{\nu}^{\dagger}\bm{\tau}\Psi_{\nu})\nonumber\\
&\hspace{-0.5cm}+2A_{2}^{2}\big[\bm{m}\Cross\partial_{x}\bm{m}+\bm{n}\Cross\partial_{x}\bm{n}\big]\bigcdot\big[-a\sin(k_{\nu,x}a)\Psi_{\nu}^{\dagger}\bm{\tau}\Psi_{\nu}+a^{2}\cos(k_{\nu,x}a)i\Psi_{\nu}^{\dagger}\bm{\tau}\partial_{x}\Psi_{\nu}\big]\nonumber\\
&\hspace{-0.5cm}+2A_{2}^{2}\big[\bm{m}\Cross\partial_{y}\bm{m}+\bm{n}\Cross\partial_{y}\bm{n}\big]\bigcdot\big[-a\sin(k_{\nu,y}a)\Psi_{\nu}^{\dagger}\bm{\tau}\Psi_{\nu}+a^{2}\cos(k_{\nu,y}a)i\Psi_{\nu}^{\dagger}\bm{\tau}\partial_{y}\Psi_{\nu}\big]\Big\}+\textrm{h.c.},\nonumber
\end{align}

\begin{align}
\label{eq:tb_expansion1_t_and_tp}
\mathcal{E}_{\textrm{tb},\nu}^{1,\textrm{intra}}\big|_{t+t'}&=\frac{1}{\mathscr{v}_{\textrm{uc}}}\int d\vec{r}\,\bigg\{\Big[A_{1}^{2}+A_{2}^{2}(\bm{m}^{2}+2\bm{m}\bigcdot\bm{n}+\bm{n}^{2})\Big]\Big[c_{+}[k_{\nu,x},k_{\nu,y}]\Psi_{\nu}^{\dagger}\Psi_{\nu}+as_{+}[k_{\nu,x},k_{\nu,y}]i\Psi_{\nu}^{\dagger}\partial_{x}\Psi_{\nu}\\
&\hspace{-1.5cm}+as_{+}[k_{\nu,x},k_{\nu,y}]i\Psi_{\nu}^{\dagger}\partial_{y}\Psi_{\nu}+a^{2}\big[\tfrac{1}{2}c_{+}[k_{\nu,x},k_{\nu,y}]\big(\Psi_{\nu}^{\dagger}\partial_{x}^{2}\Psi_{\nu}+\Psi_{\nu}^{\dagger}\partial_{y}^{2}\Psi_{\nu}\big)+c_{-}[k_{\nu,x},k_{\nu,y}]\Psi_{\nu}^{\dagger}\partial_{x}\partial_{y}\Psi_{\nu}\big]\Big]\nonumber\\
&\hspace{-1.5cm} +\tfrac{1}{2}A_{2}^{2}a^{2}\Big[c_{+}[k_{\nu,x},k_{\nu,y}](\partial_{x}[\bm{m}+\bm{n}]^{2}\Psi_{\nu}^{\dagger}\partial_{x}\Psi_{\nu}+\partial_{y}[\bm{m}+\bm{n}]^{2}\Psi_{\nu}^{\dagger}\partial_{y}\Psi_{\nu})+c_{-}[k_{\nu,x},k_{\nu,y}](\partial_{x}[\bm{m}+\bm{n}]^{2}\Psi_{\nu}^{\dagger}\partial_{y}\Psi_{\nu}\nonumber\\
&\hspace{-1.5cm}+\partial_{y}[\bm{m}+\bm{n}]^{2}\Psi_{\nu}^{\dagger}\partial_{x}\Psi_{\nu})\Big]+\tfrac{1}{2}A_{2}^{2}a^{2}\Big[c_{+}[k_{\nu,x},k_{\nu,y}]\big[(\bm{m}+\bm{n})\bigcdot\partial_{x}^{2}(\bm{m}+\bm{n})+(\bm{m}+\bm{n})\bigcdot\partial_{y}^{2}(\bm{m}+\bm{n})\big]\nonumber\\
&\hspace{-1.5cm}+2c_{-}[k_{\nu,x},k_{\nu,y}](\bm{m}+\bm{n})\bigcdot\partial_{x}\partial_{y}(\bm{m}+\bm{n})\Big]\Psi_{\nu}^{\dagger}\Psi_{\nu}+\textrm{the same integrand with }t\leftrightarrow t'\textrm{ and }\bm{n}\rightarrow -\bm{n}\bigg\}+\textrm{h.c.},\nonumber
\end{align}

\begin{align}
\label{eq:tb_expansion2_t_and_tp}
\mathcal{E}_{\textrm{tb},\nu}^{2,\textrm{intra}}\big|_{t+t'}&=\frac{1}{\mathscr{v}_{\textrm{uc}}}\int d\vec{r}\,\bigg\{A_{1}A_{2}\big(\bm{m}+\bm{n}\big)\bigcdot\Big[2c_{+}[k_{\nu,x},k_{\nu,y}]\Psi_{\nu}^{\dagger}\bm{\tau}\Psi_{\nu}+2as_{+}[k_{\nu,x},k_{\nu,y}]i\Psi_{\nu}^{\dagger}\bm{\tau}\partial_{x}\Psi_{\nu}+2as_{-}[k_{\nu,x},k_{\nu,y}]\times\\
&\hspace{-1.5cm}i\Psi_{\nu}^{\dagger}\bm{\tau}\partial_{y}\Psi_{\nu}+a^{2}c_{+}[k_{\nu,x},k_{\nu,y}]\big(\Psi_{\nu}^{\dagger}\bm{\tau}\partial_{x}^{2}\Psi_{\nu}+\Psi_{\nu}^{\dagger}\bm{\tau}\partial_{y}^{2}\Psi_{\nu}\big)+2a^{2}c_{-}[k_{\nu,x},k_{\nu,y}]\Psi_{\nu}^{\dagger}\bm{\tau}\partial_{x}\partial_{y}\Psi_{\nu}\Big]+A_{1}A_{2}a^{2}\Big[c_{+}[k_{\nu,x},k_{\nu,y}]\times\nonumber\\
&\hspace{-1.5cm}\big(\partial_{x}[\bm{m}+\bm{n}]\bigcdot(\Psi_{\nu}^{\dagger}\bm{\tau}\partial_{x}\Psi_{\nu})+\partial_{y}[\bm{m}+\bm{n}]\bigcdot(\Psi_{\nu}^{\dagger}\bm{\tau}\partial_{y}\Psi_{\nu})\big)+c_{-}[k_{\nu,x},k_{\nu,y}]\big(\partial_{x}[\bm{m}+\bm{n}]\bigcdot(\Psi_{\nu}^{\dagger}\bm{\tau}\partial_{y}\Psi_{\nu})+\partial_{y}[\bm{m}+\bm{n}]\bigcdot(\Psi_{\nu}^{\dagger}\bm{\tau}\partial_{x}\Psi_{\nu})\big)\Big]\nonumber\\
&\hspace{-1.5cm}+\tfrac{1}{2}A_{1}A_{2}a^{2}\Big[c_{+}[k_{\nu,x},k_{\nu,y}]\big(\partial_{x}^{2}+\partial_{y}^{2}\big)[\bm{m}+\bm{n}]+2c_{-}[k_{\nu,x},k_{\nu,y}]\partial_{x}\partial_{y}[\bm{m}+\bm{n}]\Big]\bigcdot\big(\Psi^{\dagger}_{\nu}\bm{\tau}\Psi_{\nu}\big)\nonumber\\
&\hspace{-1.5cm}-A_{2}^{2}a\Big[s_{+}[k_{\nu,x},k_{\nu,y}]\big(\bm{m}+\bm{n}\big)\Cross\partial_{x}[\bm{m}+\bm{n}]+s_{-}[k_{\nu,x},k_{\nu,y}]\big(\bm{m}+\bm{n}\big)\Cross\partial_{y}[\bm{m}+\bm{n}]\Big]\bigcdot\big(\Psi^{\dagger}_{\nu}\bm{\tau}\Psi_{\nu}\big)\nonumber\\
&\hspace{-1.5cm}+A_{2}^{2}a^{2}\Big[c_{+}[k_{\nu,x},k_{\nu,y}]\big[\big((\bm{m}+\bm{n})\Cross\partial_{x}[\bm{m}+\bm{n}]\big)\bigcdot(i\Psi_{\nu}^{\dagger}\bm{\tau}\partial_{x}\Psi_{\nu})+\big((\bm{m}+\bm{n})\Cross\partial_{y}[\bm{m}+\bm{n}]\big)\bigcdot(i\Psi_{\nu}^{\dagger}\bm{\tau}\partial_{y}\Psi_{\nu})\big]\nonumber\\
&\hspace{-1.5cm}+c_{-}[k_{\nu,x},k_{\nu,y}]\big[\big((\bm{m}+\bm{n})\Cross\partial_{x}[\bm{m}+\bm{n}]\big)\bigcdot(i\Psi_{\nu}^{\dagger}\bm{\tau}\partial_{y}\Psi_{\nu})+\big((\bm{m}+\bm{n})\Cross\partial_{y}[\bm{m}+\bm{n}]\big)\bigcdot(i\Psi_{\nu}^{\dagger}\bm{\tau}\partial_{x}\Psi_{\nu})\big]\nonumber\\
&\hspace{-1.5cm}+\textrm{the same integrand with }t\leftrightarrow t'\textrm{ and }\bm{n}\rightarrow -\bm{n}\bigg\}+\textrm{h.c.},\nonumber
\end{align}
\end{widetext}
where $c_{\pm}[k_{x},k_{y}]\equiv t\cos[(k_{x}+k_{y})a]\pm t'\cos[(k_{x}-k_{y})a]$ and $s_{\pm}[k_{x},k_{y}]\equiv t\sin[(k_{x}+k_{y})a]\pm t'\cos[(k_{x}-k_{y})a]$ are auxiliary trigonometric functions depending on the valley wavevector $\vec{k}_{\nu}$. Furthermore, we have also derived an analogous long-wavelength expansion of the interlattice contribution to the energy functional, which we omit here since it does not generate any new relevant coupling terms.

Similar considerations apply to the mean-field path-integral kinetic Lagrangian for fermions, see Eq.~\eqref{eq:kin_term}, so that the following expression is valid irrespective of the choice of valley/band minimum in the low-energy long-wavelength limit:
\begin{widetext}
\begin{align}
\label{eq:kin_Lagrangian}
\mathcal{L}_{\textrm{kin}}&=\frac{\hbar}{\mathscr{v}_{\textrm{uc}}}\int d\vec{r}\,\Big[(A_{1}^{2}+A_{2}^{2}\,\mathfrak{s}^{2})(\Psi^{\dagger}_{\nu}\partial_{0}\Psi_{\nu}-\partial_{0}\Psi^{\dagger}_{\nu}\Psi_{\nu})+2A_{1}A_{2}\bm{m}\bigcdot(\Psi^{\dagger}_{\nu}\bm{\tau}\partial_{0}\Psi_{\nu}-\partial_{0}\Psi^{\dagger}_{\nu}\bm{\tau}\Psi_{\nu})\\
&\hspace{2.2cm}+2iA_{2}^{2}(\bm{m}\Cross\partial_{0}\bm{m}+\bm{n}\Cross\partial_{0}\bm{n})\bigcdot(\Psi^{\dagger}_{\nu}\bm{\tau}\Psi_{\nu})\Big].\nonumber
\end{align}
\end{widetext}

In the main text we have specified these long-wavelength expansions at the valley points of the tight-binding model considered, namely the $\bm{\Gamma}$ and $\bm{Z}$ points of the Brillouin zone. These energy minima are characterized by the wavevectors $\vec{k}_{\Gamma}=(0,0,0)$ and $\vec{k}_{Z}=(0,0,\tfrac{\pi}{c})$, respectively, both leading to the general expression~\eqref{eq:EuclidLag_AM} for the effective Euclidean Lagrangian describing the physics of the itinerant carriers. We note that Eq.~\eqref{eq:EuclidLag_AM} has been obtained by disregarding surface terms as well as terms being O(3) in spatial derivatives and total magnetization. Furthermore, the coupling constants of our theory at the $\bm{\Gamma}$ point are defined in terms of the microscopic parameters of the model as:
\begin{widetext}
\begin{align}
\label{eq:couplings_AFM}
&g_{0}\equiv\frac{\mathcal{V}}{\mathscr{v}_{\textrm{uc}}}\hbar\big(A_{1}^{2}+A_{2}^{2}\mathfrak{s}^{2}\big),\hspace{0.2cm}g_{0}'\equiv2\frac{\mathcal{V}}{\mathscr{v}_{\textrm{uc}}}\hbar A_{1}A_{2},\hspace{0.2cm}g_{0}''\equiv2\frac{\mathcal{V}}{\mathscr{v}_{\textrm{uc}}}\hbar A_{2}^{2},\\
&g_{1}^{m}=4\frac{\mathcal{V}}{\mathscr{v}_{\textrm{uc}}}\big(t_{z}+2t_{0}+t+t'+4t^{AB}\big)A_{2}^{2},\hspace{0.2cm}g_{1}^{z}=\frac{\mathcal{V}}{\mathscr{v}_{\textrm{uc}}}(t^{AB}-t_{z})c^{2}A_{2}^{2},\hspace{0.2cm}g_{1}^{xy}=\frac{\mathcal{V}}{\mathscr{v}_{\textrm{uc}}}(t^{AB}-t_{0}-t-t')a^{2}A_{2}^{2},\nonumber\\
&g_{1}^{\textrm{AM}}=2\frac{\mathcal{V}}{\mathscr{v}_{\textrm{uc}}}(t-t')a^{2}A_{2}^{2},\hspace{0.2cm}g_{2}^{m}=8\frac{\mathcal{V}}{\mathscr{v}_{\textrm{uc}}}(t_{z}+2t_{0}+t+t'+4t^{AB})A_{1}A_{2},\hspace{0.2cm}g_{2}^{\textrm{AM}}=4\frac{\mathcal{V}}{\mathscr{v}_{\textrm{uc}}}(t-t')a^{2}A_{1}A_{2},\nonumber\\
&g_{4}^{xy}=2\frac{\mathcal{V}}{\mathscr{v}_{\textrm{uc}}}(t_{0}+t+t'-t^{AB})a^{2}A_{2}^{2},\hspace{0.2cm}g_{4}^{\textrm{AM}}=2\frac{\mathcal{V}}{\mathscr{v}_{\textrm{uc}}}(t-t')a^{2}A_{2}^{2},\hspace{0.2cm}g_{4}^{z}=2\frac{\mathcal{V}}{\mathscr{v}_{\textrm{uc}}}(t_{z}-t^{AB})c^{2}A_{2}^{2},\nonumber\\
&\frac{\hbar g_{0}}{m_{\parallel}}=2\frac{\mathcal{V}}{\mathscr{v}_{\textrm{uc}}}\big[t^{AB}(A_{2}^{2}\mathfrak{s}^{2}-A_{1}^{2})-(t_{0}+t+t')(A_{1}^{2}+A_{2}^{2}\mathfrak{s}^{2})\big]a^{2},\hspace{0.2cm}\frac{\hbar g_{0}}{m_{z}}=2\frac{\mathcal{V}}{\mathscr{v}_{\textrm{uc}}}\big[t^{AB}(A_{2}^{2}\mathfrak{s}^{2}-A_{1}^{2})-t_{z}(A_{1}^{2}+A_{2}^{2}\mathfrak{s}^{2})\big]c^{2},\nonumber\\
&\frac{\hbar g_{0}}{m_{xy}^{s}}=4\frac{\mathcal{V}}{\mathscr{v}_{\textrm{uc}}}(t'-t)a^{2}A_{1}A_{2},\hspace{0.2cm}\frac{\hbar g_{0}}{m_{\parallel}^{s}}=4\frac{\mathcal{V}}{\mathscr{v}_{\textrm{uc}}}(t_{0}+t+t'+t^{AB})a^{2}A_{1}A_{2},\hspace{0.2cm}\frac{\hbar g_{0}}{m_{z}^{s}}=4\frac{\mathcal{V}}{\mathscr{v}_{\textrm{uc}}}(t_{z}+t^{AB})c^{2}A_{1}A_{2}.\nonumber
\end{align}
\end{widetext}
Here, $\mathcal{V}$ denotes the total volume of the magnet, which appears in the above expressions for the coupling constants due to the normalization condition $\Psi\rightarrow\tfrac{1}{\sqrt{\mathcal{V}}}\Psi$ for the Fermi field. We note that these same expressions do apply at the $\bm{Z}$-point with account of the sign flip $t_{z}\rightarrow -t_{z}$.

\section{Intermediate expressions}

\label{AppA3}

The following identity has been used in the derivation of the long-wavelength energy functional~\eqref{eq:energy_itinerant} for the itinerant carriers in terms of the thermodynamic variables of the system:

\begin{widetext}
\begin{align}
\label{eq:ident2}
&\bm{\mathcal{A}}_{\alpha}\bigcdot(i\partial_{\alpha}\Psi^{\dagger}\bm{\tau}\Psi-i\Psi^{\dagger}\bm{\tau}\partial_{\alpha}\Psi)=\tfrac{4m_{z}}{\hbar^{2}}\big[\bm{\mathcal{A}}_{z}\bigcdot\bm{J}_{z}+\tfrac{\hbar}{2g_{0}}\rho\bm{\mathcal{A}}_{z}^{2}\big]+\tfrac{2m_{\parallel}}{\hbar}\sum_{\kappa=x,y}\Big[\tfrac{2}{\hbar}\bm{J}_{\kappa}\bigcdot\bm{\mathcal{A}}_{\kappa}+\tfrac{1}{g_{0}}\rho\bm{\mathcal{A}}_{\kappa}^{2}\\
&\hspace{1cm}+\tfrac{2}{\hbar}\tfrac{1}{(m_{xy}^{s}/m_{\parallel})^{2}-\mathfrak{s}^{2}}(\bm{n}\bigcdot\bm{\mathcal{A}}_{\kappa})(\bm{n}\bigcdot\bm{J}_{\kappa})+\tfrac{1}{g_{0}}\tfrac{\rho}{(m_{xy}^{s}/m_{\parallel})^{2}-\mathfrak{s}^{2}}(\bm{n}\bigcdot\bm{\mathcal{A}}_{\kappa})^{2}\Big]-\tfrac{2}{\hbar}\tfrac{m_{\parallel}}{m_{xy}^{s}}\sum_{\kappa,\beta=x,y}\Big[\sigma_{x}|_{\kappa\beta}\bm{\mathcal{A}}_{\kappa}\bigcdot(\bm{n}\Cross\partial_{\beta}\bm{s})\Big]\nonumber\\
&\hspace{1cm}-\tfrac{2}{\hbar}\tfrac{m_{xy}^{s}}{(m_{xy}^{s}/m_{\parallel})^{2}-\mathfrak{s}^{2}}\sum_{\kappa,\beta=x,y}\Big[\sigma_{x}|_{\kappa\beta}\big(j_{\beta}+\tfrac{2}{\hbar g_{0}}\bm{\mathcal{A}}_{\beta}\bigcdot\bm{s}\big)(\bm{n}\bigcdot\bm{\mathcal{A}}_{\kappa})\Big]\nonumber.
\end{align}
\end{widetext}

The functional derivatives of the energy functional $\mathcal{E}_{\textrm{ST}}$ with respect to the components of the N\'{e}el order and the macroscopic spin density are given by

\begin{widetext}
\begin{align}
\label{eq:func_deriv}
\frac{\delta\mathcal{E}_{\textrm{ST}}}{\delta n^{\beta}(\vec{r}\,')}&=-\tfrac{4m_{\parallel}g_{4}^{xy}}{\hbar^{2}}\epsilon_{\alpha\beta\gamma}J_{\kappa}^{\alpha}(\vec{r}\,')\partial_{\kappa}n^{\gamma}(\vec{r}\,')+\tfrac{4m_{\parallel}g_{4}^{xy}}{\hbar^{2}}\epsilon_{\alpha\beta'\beta}\partial_{\kappa}\big[J_{\kappa}^{\alpha}n^{\beta'}\big](\vec{r}\,')+\tfrac{4m_{\parallel}g_{4}^{\textrm{AM}}}{\hbar^{2}}\epsilon_{\alpha\beta'\beta}\sigma_{x}|_{\kappa\sigma}\partial_{\sigma}\big[J_{\kappa}^{\alpha}m^{\beta'}\big](\vec{r}\,'),\\
\frac{\delta\mathcal{E}_{\textrm{ST}}}{\delta m^{\beta}(\vec{r}\,')}&=-\tfrac{4m_{\parallel}g_{4}^{\textrm{AM}}}{\hbar^{2}}\epsilon_{\alpha\beta\gamma}\sigma_{x}|_{\kappa\sigma}J_{\kappa}^{\alpha}(\vec{r}\,')\partial_{\sigma}n^{\gamma}(\vec{r}\,').\nonumber
\end{align}
\end{widetext}


\begin{thebibliography}{99}
\bibitem{FN1} In other words, these off-diagonal terms cost more energy due to the presence of phases of the form $e^{i(\vec{k}_{\nu}-\vec{k}_{\bar{\nu}})\cdot\vec{r}}\neq0$, and therefore are neglected in the low-energy long-wavelength limit.
\end{thebibliography}
\end{document}